\documentclass[draftclsnofoot,onecolumn]{IEEEtran}
\usepackage{amsmath,amssymb,amsfonts}
\usepackage{algorithmic}
\usepackage{graphicx}
\usepackage{textcomp}
\usepackage{xcolor}
\usepackage{url}
\usepackage{ifthen}
\usepackage{multicol}
\usepackage{latexsym}
\usepackage{wrapfig}
\usepackage{blindtext}
\usepackage{bm}
\usepackage{lipsum}
\usepackage{cuted}
\usepackage{amsthm}
\usepackage{ascmac}
\usepackage{lipsum}
\usepackage{mathtools}
\usepackage{cuted}
\usepackage{enumitem}
\usepackage{xcolor}
\usepackage{subcaption}
\usepackage{comment}
\usepackage{cite}
\usepackage{dsfont}
\usepackage[font=footnotesize]{caption}
\usepackage{colortbl}
\usepackage{hyperref}
\usepackage[normalem]{ulem}
\hypersetup{
    colorlinks=true,  
    linkcolor= blue,  
    citecolor=blue,  
    urlcolor=black  
}
 \newtheorem{Theorem}{Theorem}
\newtheorem{Definition}{Definition}

\newtheorem{Lemma}{Lemma}
\newtheorem{Example}{Example}

\newtheorem{Setup}{Setup}

\def\V{{\mathbb{V}}}
\def\Var{{\rm Var}}
\def\MMSE{{\rm MMSE}}
\def\E{{\mathbb{E}}}

\def\M{\mathcal{M}}

\def\A{\mathcal{A}}
\def\B{\mathcal{B}}

\def\S{\mathcal{S}}

\def\Si{\mathcal{S}_i}

\def\P{\mathcal{P}}

\def\Pi{\mathcal{P}_i}

\def\I{\mathcal{I}}

\def\X{\mathcal{X}}
\def\Y{\mathcal{Y}}

\ifCLASSINFOpdf

\else
  
\fi

\begin{document}
\sloppy
\allowdisplaybreaks[1]
\title{Privacy-Constrained Function Computation\\in Distributed Databases}

\author{Vamoua~Yachongka,
        R\'emi~A.~Chou
\thanks{Vamoua Yachongka is with the Faculty of Engineering, The University of the Ryukyus, 1 Senbaru, Nishihara, Nakagami, Okinawa, 903-0129 Japan. R\'emi A. Chou is with the Department of Computer Science and Engineering, The University of Texas at Arlington, Arlington, TX, 76010 USA. E-mails: (vamoua@ie.u-ryukyu.ac.jp, remi.chou@uta.edu). A preliminary version of this paper was presented at the 2025 IEEE International Symposium on Information Theory (ISIT 2025) \cite{vr-isit2025}. This work was supported in part by NSF grant CCF-2425371.}
}

\markboth{}
{Shell \MakeLowercase{\textit{et al.}}: Bare Demo of IEEEtran.cls for IEEE Journals}

\maketitle

\begin{abstract}
Consider distributed databases that contain both sensitive (private) and non-sensitive (public) attributes. A user seeks to compute a function of public attributes and side information by interacting with the databases and downloading their responses. In this paper, we study the trade-offs among download cost, information leakage about the private attributes, and distortion between the computed function and its reconstruction, referred to as the capacity region. Our main contributions are as follows. For a single database, we provide a complete characterization of the capacity region for general discrete sources and characterize the capacity region of two different setups for Gaussian sources when the computed function is defined as the sum of the public attribute and side information. For distributed databases, we establish inner and outer bounds on the capacity region of discrete sources and we characterize the capacity region when the databases
and the side information are mutually independent. As a special instance, we study lossless function computation, and results analogous to the lossy case are obtained.
\end{abstract}

\IEEEpeerreviewmaketitle

\section{Introduction} \label{sect1}

The advancement of the internet and Internet of Things (IoT) technologies, such as real-time remote monitoring devices like smartwatches and heart rate sensors, has led to the generation and collection of vast amounts of data, which are commonly stored in databases (DBs) like cloud storage due to limited internal memory. While this enables convenient access, it also raises significant privacy concerns, particularly when handling sensitive data \cite{Ali-2021}.
Public attributes such as interaction logs between the system and users can be encoded and shared with the system administrator, referred to as the decoder in this paper, for purposes such as improving system usability. On the other hand, private attributes associated with personal data must be safeguarded to preserve user privacy. However, correlations may still allow the decoder to infer private attributes from the shared data and any available side information. Therefore, it is crucial to design coding schemes that maximize the system usability and limit the decoder's ability to infer private attributes.

Rate-distortion theory has been used to analyze the optimal utility-privacy trade-offs in DBs in \cite{yamamoto-1983, Sankar-2013}. Specifically, \cite{yamamoto-1983} considers a scenario where a source generates a private and a public attribute. The encoder transmits an encoded message of these attributes to the decoder, which aims to reconstruct the public attribute within a specified distortion level. At the same time, the uncertainty of the private attribute for the decoder should be as large as possible. The scenario is later extended in \cite{Sankar-2013} to incorporate multiple private and public attributes as well as side information at the decoder. In these studies,  private and public attributes are stored in the same DB and the decoder's objective is to reconstruct the public attributes. However, compared to a single database, distributed DBs have several advantages, particularly in terms of scalability and security risk diversification~\cite{Domdouzis2021}. Furthermore, the decoder might only be interested in learning the value of a function of the attributes observed at local sites rather   than the entire collection of observed attributes \cite{alphan-2023}.

In this paper, therefore, we consider a system where the attributes are stored across multiple DBs in a distributed manner and the task of the decoder is to compute a function of public attributes and side information. Specifically, the decoder interacts with the DBs to download encoded responses derived from their public attributes and combines them with the side information to compute the function. Our goal is to characterize the optimal trade-offs among the download cost of responses, the information leakage about the private attributes that the decoder may infer, and the distortion between the target function and its reconstruction for both discrete and Gaussian sources. Throughout this paper, the optimal trade-offs, download cost, and information leakage are referred to as the \emph{capacity region}, \emph{communication rate}, and \emph{leakage rate}, respectively.

\subsection{Related Work}
We now review prior works relevant to our study. In particular, we begin by reviewing results on function computation with only communication constraints, then discuss function computation under both communication and privacy constraints, and finally highlight the literature most closely related to this work.

\subsubsection{Function Computation with Communication Constraints Only}
Lossless function computation with a communication constraint is studied in \cite{orlitsky-2001}, and the optimal communication rate is characterized when the decoder has side information. This framework is later extended in \cite{milad-arxiv} to study distributed function computation, where inner and outer bounds on the capacity region of communication rates are derived. Further generalization is presented in \cite{milad2016}, which investigates distributed function computation over a rooted-directed tree, where a node receives messages from its children and communicates with its parent to help in computing a function at the root. Several works have investigated the capacity region for binary sources with specific functions such as addition \cite{km-1979,ah-1983,chandra-2020,milad-2020,tharindu-2022,guang-2024,zhang-2024}. Scenarios where the computed function is not necessarily symbol-wise and the source may have memory are explored in \cite{kuzuoka-2015,kuzuoka-2017,wata-2020}. Unlike these works \cite{orlitsky-2001,milad-arxiv,milad2016,km-1979,ah-1983,chandra-2020,milad-2020,tharindu-2022,guang-2024,zhang-2024,kuzuoka-2015,kuzuoka-2017,wata-2020}, our study considers private attributes and imposes a privacy constraint to minimize the leakage of these private attributes.

\subsubsection{Function Computation with Communication and Privacy Constraints}
Several works considered function computation subject to privacy constraints. In \cite{deepesh-2016} and \cite{deepesh-2020}, a model where all the involved users are curious about each other's data and have access to  common randomness is studied. The common randomness is essential in achieving negligible privacy leakage as it is used to conceal information  from other users. In  the absence of common randomness, vanishing leakage need not be achievable \cite{tu2019,lu-2022}. The model in \cite{tu2019,lu-2022} is further generalized in \cite{chou2022,chou-isit2024}, which examine multi-user scenarios involving colluding users. In contrast to these works, where the privacy constraint is defined with respect to the sequences observed at the encoders (corresponding to public attributes in our model settings), in this paper, the privacy constraint is measured based on private attributes that are hidden from the encoders.

\subsubsection{Most Related Literature}
The studies most closely related to our work are  \cite{Sankar-2013,Yamamoto-fc, milad-arxiv,tu2019,gunlu-en-d}. As alluded to earlier, we extend the model in \cite{Sankar-2013} by incorporating distributed DBs and function computation at the decoder. Distributed lossless function computation \cite{milad-arxiv} and lossy function computation for the Wyner-Ziv model \cite{Yamamoto-fc} differ from our work in that they do not account for private attributes, and no privacy constraint is imposed in their problem formulation. The works \cite{tu2019,gunlu-en-d} {derive inner and outer bounds for distributed} lossless and lossy function computation under privacy constraints. Specifically, \cite{tu2019} considers a model with  {two public attributes, where each encoder observes one public attribute}, and \cite{gunlu-en-d} extends the work \cite{tu2019} to include {one private attribute}. Moreover, \cite{gunlu-en-d} assumes conditional independence between the encoder observation (the public attribute) and side information given the private attribute, similar to the key-generation problem with a privacy constraint \cite{gunlu-et-al2018,vamoua-tifs,va-remi-tifs}. However, this conditional independence assumption may not fully capture the correlation structure arising in database systems, where the underlying random variables are commonly modeled as jointly correlated \cite{Sankar-2013}.

In this paper, we deal with both lossless and lossy scenarios like \cite{tu2019,gunlu-en-d}. However, we go beyond the model in \cite{gunlu-en-d} by incorporating multiple private and public attributes and relaxing the conditional independence between public attributes and side information. Furthermore, while \cite{gunlu-en-d} focuses on a two-encoder model with discrete sources, we study a multiple-encoder (more than two) model and investigate the capacity region for Gaussian sources, which requires proof techniques distinct from those used in the discrete case. While function computation without privacy constraints over Gaussian channels has been considered in the literature, e.g., \cite{nazer-2007,Jeon-2014,goldenbuam,frey-2025}, to the best of our knowledge, distributed function computation with privacy constraints in Gaussian settings has not yet been investigated.

\subsection{Main Challenges and Contributions}
In this subsection, we highlight some key challenges encountered in establishing our main results for both single and distributed DB settings, followed by a summary of the main findings presented in this paper.
\subsubsection{Main Challenges}
For a single DB, the main difficulty lies in proving the converse of the capacity region for Gaussian sources in lossy function computation, where the computed function is defined as the sum of the public attribute and side information. We consider two different setups. In the first setup, the private attribute and side information are noisy observations of the public attribute as in \cite[Ex.~2]{Sankar-2013}. In the second setup, the public attribute and the side information are noisy versions of the private attribute, motivated by the hidden private-data scenario in \cite{amirraza-2024}. While existing works on Gaussian multiterminal systems with privacy constraints \cite{villard-2013,kitti-et-al2016,naghibis2015,va-tifs-2025} rely heavily on the conditional entropy power inequality (EPI) \cite{bergmans1974} to derive outer bounds, this method does not seem easily applicable to our setups. To address this challenge, we employ a technique based on the maximum entropy property and the monotonicity of the logarithmic function to derive the outer bound for the first setup. For the second setup, we use a Fisher-information argument \cite{ekrem-2014} to parametrize the minimum mean squared error (MMSE) induced by the auxiliary variable and exploit its relation to Fisher information.

For distributed DBs, compared to \cite{tu2019,gunlu-en-d}, as our model extends beyond the two-encoder scenario, the proof becomes more involved. Specifically, in the achievability part, an extra step that exploits the contra-polymatroid structure of the achievable rates is essential. The step demonstrates that showing the achievability of a corner point on the dominant face suffices to establish the achievability proof. In the converse part, relaxing the conditional independence assumption leads to a distinct choice of the auxiliary random variable when deriving the outer bound. Additionally, lossless function computation requires different bounding techniques to handle the exact reconstruction constraint. Finally, because correlations among the auxiliary variables make it difficult to apply the support lemma \cite{GK}, we establish a variable-replacement method to derive their cardinality bounds.

\subsubsection{Main Contributions}
The main contributions of this paper are summarized as~follows.
\begin{itemize}
\item We characterize the capacity region between the communication and leakage rates in a single DB for lossless function computation with general discrete sources.
We note that at zero Hamming distortion, the lossy criterion requires only that
the expected fraction of incorrectly reconstructed function values
vanishes. Hence, we prove lossless achievability separately to establish the
stronger condition that the probability of any reconstruction error
in the entire block vanishes.
\item For a single DB, we further characterize the capacity region of communication rate, leakage rate, and distortion for lossy function computation, where the decoder reconstructs only a distorted version of the target function. This characterization is then applied to derive the capacity regions of two different setups for Gaussian sources when the target function is defined as the sum of the public attribute and side information.
\item For both lossless and lossy function computations, we establish inner and outer bounds on the capacity region for distributed DBs and characterize the capacity region when {the DBs and side information are mutually independent}.
\end{itemize}

Special cases of our results recover several existing results. Specifically, in a special case of discrete sources and lossless function computation for a single database, our result recovers \cite[Th.~1]{orlitsky-2001} when there is no private attribute and one public attribute in the database. Furthermore, for lossy function computation, our result recovers  \cite[Th.~2]{Sankar-2013} in a single database when the target function equals the public attributes. Our result also reduces to \cite[Th.~1]{Yamamoto-fc} when there is no private attribute and one public attribute. For Gaussian sources, when the target function is the public attribute and the leakage metric is replaced by unconditional mutual information, the capacity result of the first setup matches the capacity region of \cite[Case~(ii) in Ex.~2]{Sankar-2013}. Finally, when the function is the public attribute and there is no side information, the result of the first setup reduces to \cite[Case~(i) in Ex.~2]{Sankar-2013}, and to \cite[Remark~6]{wyner-ziv} without the leakage rate.

The remainder of the paper is organized as follows. Section~\ref{sect2} describes the   problem statement. Sections \ref{sect3} and~\ref{sect4} present our main results for a single DB and distributed DBs, respectively. The detailed proofs are deferred to the appendices. Finally, Section~\ref{sect5} provides concluding remarks.

\section{Problem Statement} \label{sect2}
\emph{Notation}: $\mathbb{R}_+$ represents the set of non-negative real numbers. For any $p,q \in \mathbb{R}$, define $[p:q] \triangleq [\lfloor p \rfloor,\lceil q \rceil] \cap \mathbb{N}$. The uppercase $X$ and lowercase $x$ represent a random variable and its realization, respectively. The calligraphic letter $\mathcal{A}$ denotes a set and the $n$-ary Cartesian power of it is denoted by $\mathcal{A}^n$. $X_{\mathcal{A}} \triangleq (X_i)_{i \in \mathcal{A}}$  refers to a collection of random variables and takes values in $\mathcal{X}_{\A} \triangleq \bigtimes_{i \in \mathcal{A}} \mathcal{X}_i$. Similarly, $X^n_{\mathcal{A}} \triangleq (X^n_i)_{i \in \mathcal{A}}$ represents a collection of vector random variables and takes values in $\mathcal{X}^n_{\A} \triangleq \bigtimes_{i \in \mathcal{A}} \mathcal{X}^n_i$. For an integer $t~(\le n)$, the subscript $X_{\mathcal{A},t}$ indicates the collection of random variables at position $t$ in $X^n_{\mathcal{A}}$, while the superscript $X^t_{\mathcal{A}}$ denotes the partial sequence $(X_{\mathcal{A},i})_{i \in [1:t]}$. {We define $Y^{n \backslash t} \triangleq (Y_1, \ldots, Y_{t-1}, Y_{t+1}, \ldots, Y_n)$}, and let $\mathcal{N}(0,\sigma^2)$ denote the Gaussian distribution with zero mean and variance $\sigma^2$. The conditional variance of $X$ given $U$ is denoted by $\mathrm{Var}[X|U]$. {$\log(\cdot)$ represents the base-$2$ logarithm}. Finally, $\mathds{1}\{\mathcal{E}\}$ denotes the indicator function, which equals $1$ if the event $\mathcal{E}$ is true and $0$ otherwise.

Let $\I \triangleq [1:\kappa]$ be the index set of DBs. For $i \in \I$,   $\S_i$ and $\P_i$ denote the sets of indices of private and public attributes in the $i$-th database, respectively, and satisfy $\S_i\cap\P_i = \emptyset$ with $\bigcup_{i=1}^{\kappa}\S_i = \S$ and $\bigcup_{i=1}^{\kappa}\P_i = \P$, where the sets $\S$ and $\P$ denote the indices of private and public attributes across all databases, respectively.
\begin{figure}[t]
\centering
   \includegraphics[scale=0.7]{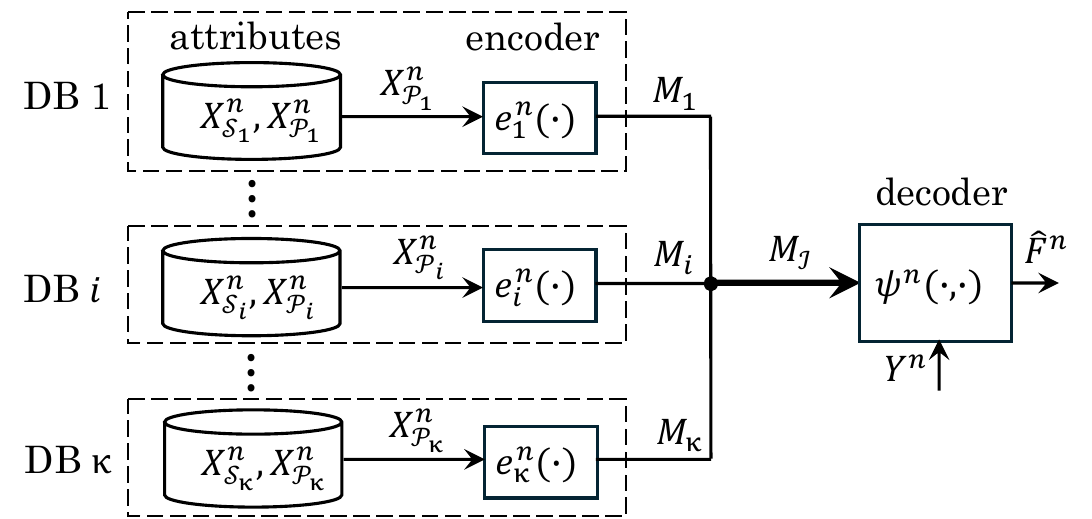}
   \caption{Our system model consists of $\kappa$ DBs and a decoder. The $i$-th DB stores the private and public attributes $X^n_{\S_i}$ and $X^n_{\P_i}$, which are correlated with the side information $Y^n$. Each DB is equipped with an encoder that produces a response $M_i$ based on $X^n_{\P_i}$. The decoder retrieves responses from the DBs and uses them together with $Y^n$ to estimate $F^n$ as $\hat{F}^n$.}
   \label{model-1}
\end{figure}
The vector random variables $X^n_{\P}$, $X^n_{\S}$, $Y^n$ are independent and identically distributed (i.i.d.)~copies of $X_{\P}$, $X_{\S}$, $Y$, respectively, which take values from the sets $\mathcal{X}_{\S}$, $\mathcal{X}_{\P}$, and $\mathcal{Y}$, and are i.i.d. generated according to the joint distribution
\begin{align}
P_{X_{\P}X_{\S}Y} \triangleq P_{X_{\S}}P_{X_{\P}|X_{\S}}P_{Y|X_{\S}X_{\P}}. \label{joint-dist}
\end{align}
Our system model is depicted in Fig. \ref{model-1}.  The goal of the decoder is to compute a function $ f^n(X^n_{\P},Y^n) \triangleq (f(X_{\P,t},Y_t))_{t \in [1:n]} \in \mathcal{F}^n $, where $X^n_{\P}$ and $Y^n$ represent the public attributes and side information. For brevity, we write $f^n(X^n_{\P},Y^n)$ as $F^n \triangleq (F_t)_{t \in [1:n]}$.

\begin{Definition} A $((2^{nR_i}\!)_{i \in \I},n)$-computation scheme consists~of
\begin{itemize}
    \item Response alphabets $\M_i \triangleq [1:2^{nR_i}]$, for $i \in \I$;
    \item Stochastic encoding functions $e^n_i: \mathcal{X}^n_{\P_i} \rightarrow \M_i$, for $i \in \I$;
    \item A decoding function $\psi: \M_{\I}\times\mathcal{Y}^n \rightarrow \mathcal{F}^n$;
\end{itemize}
and operates as follows:
\begin{itemize}
    \item Encoder $i \in \I$ generates $M_i \triangleq e^n_i(X^n_{\P_i})$ and sends it to the decoder;
    \item The decoder forms an estimate $\hat{F}^n \triangleq \psi^n(M_{\I},Y^n)$ of $F^n$.
\end{itemize}
\end{Definition}

For finite source alphabets, the distortion incurred between $F^n$ and its estimate $\hat{F}^n$ is evaluated as
\vspace{-2mm}
\begin{align}
    d^n(F^n,\hat{F}^n) \triangleq \frac{1}{n}\sum_{t=1}^nd(F_t,\hat{F}_t), \label{distor-sum}
\end{align}
where $d: \mathcal{F}\times\mathcal{F} \rightarrow [0,D_{\max}]$ {with $D_{\max} < \infty$} is a per-letter distortion measure, and the reconstruction alphabet $\mathcal{F}$ is assumed to be finite.

\begin{Definition} \label{def2} A tuple of communication rates, leakage rate, and distortion $((R_i)_{i \in \I}, L,D) \in \mathbb{R}^{\kappa+2}_+$ is  achievable if for $\delta > 0$ and large enough $n$, there exists a sequence of $((2^{nR_i}\!)_{i \in \I},n)$-computation schemes such that
\begin{align}
    \log|\mathcal{M}_i| &\le n(R_i + \delta),~~\forall i \in \I, \label{rate-def2} \\
    I(X^n_{\S};M_{\I}|Y^n) &\le n(L + \delta), \label{pri-def2} \\
    \mathbb{E}[d^n(F^n,\hat{F}^n)] &\le D + \delta. \label{dis-def2}
\end{align}
The capacity region $\mathcal{R}_D$ is defined as the closure of the set of all achievable rate tuples.
\end{Definition}

Equations \eqref{rate-def2}, \eqref{pri-def2}, and \eqref{dis-def2} represent the constraints of the communication rate for each response, the leakage rate, and the distortion. Note that \eqref{pri-def2} quantifies the information leaked about the entire set of private attributes $X^n_{\S}$ to the decoder, which receives a set of responses $M_{\I}$ from DBs and observes side information $Y^n$.

For lossless function computation, the constraint \eqref{dis-def2} in Definition~\ref{def2} is replaced by
\begin{align}
    \mathbb{P}[F^n \neq \hat{F}^n] &\le \delta. \label{error-def1}
\end{align}
Moreover, we define $\mathcal{R}$ as the capacity region for all achievable rate tuples $((R_i)_{i \in \I},L) \in \mathbb{R}^{\kappa + 1}_+$ for this case.

\section{Single Database} \label{sect3}
In this section, we examine a scenario in which all sensitive and public attributes are stored in a single DB, i.e., $\kappa = 1$. We first present the capacity result for  lossless function computation in Section \ref{sect-3a} and then discuss the lossy case in Section \ref{sect-3b}.

\subsection{Lossless Function Computation} \label{sect-3a}
\begin{Theorem} \label{th1} Let $\Tilde{\Omega}$ be the set of probability distributions $P_{UX_{\P}X_{\S}Y}$ over $\mathcal{U}\times \X_{\P}\times \X_{\S}\times \Y$ such that $P_{UX_{\P}X_{\S}Y} = P_{X_{\P}X_{\S}Y}P_{U|X_{\P}}$ and $H(F|U,Y) = 0$, where $|\mathcal{U}| \le |\mathcal{X}_{\P}| + 2$. The capacity region~is
\begin{align}
\mathcal{R} = \bigcup_{P_{UX_{\P}X_{\S}Y} \in \Tilde{\Omega}}\Big\{(R,L) \in \mathbb{R}^2_+: 
    R \ge I(X_{\P};U|Y),~~
    L \ge I(X_{\S};U|Y)\Big\}. \label{th1-ex1}
\end{align}
\end{Theorem}
\begin{proof}
The achievability of Theorem \ref{th1} is obtained by setting $ |\I| = 1$ and $ Q = \emptyset$ in Theorem \ref{inner-bound}. The converse follows by applying the same argument as in the proof of Theorem \ref{outer-bound} with the auxiliary random variable defined as $U_t = (M, X^{t-1}_{\S}, Y^{n \backslash t})$ for $t \in [1:n]$, where $M$ denotes the response generated from the public  attributes $X^n_{\P}$. The cardinality bound follows by the support lemma \cite[Appx.~C]{GK}.
\end{proof}

Next, we provide three examples for Theorem~\ref{th1}, when $\P = \{1\}$ and $\S= \{s\}$, which corresponds to a setting where there is only one private attribute $X_s$ and one public attribute $X_1$ in the DB. For simplicity, we write the public attribute $X_1$ as $X$ in the rest of this section. 
In this case, the capacity expression in \eqref{th1-ex1} reduces to
\begin{align}
\mathcal{R}_o = \bigcup_{P_{UXX_sY} \in \Tilde{\Omega}_o}\Big\{(R,L) \in \mathbb{R}^2_+: 
    R \ge I(X;U|Y),~~
    L \ge I(X_s;U|Y)\Big\}, \label{ex1-eq111}
\end{align}
where $\Tilde{\Omega}_o$ is the set of probability distributions $P_{UXX_sY}$ over $\mathcal{U}\times \X\times \X_s\times \Y$ such that $P_{UXX_sY} = P_{XX_sY}P_{U|X}$ and $H(F|U,Y) = 0$, where $|\mathcal{U}| \le |\X| + 2$. Next, we  rewrite \eqref{ex1-eq111}  in Lemma \ref{lemma-1111} using a conditional graph entropy, whose definition is reviewed below.

\begin{Definition}[\hspace{-0.1mm}\cite{orlitsky-2001}] \label{def-3} The definitions of conditional characteristic graphs, independent sets and maximal independent sets, and conditional graph entropy are provided as follows:
\begin{enumerate}
\item Let $(X,Y) \sim P_{XY}$ represent a pair of random variables and $f(X,Y)$ be a given function, the conditional characteristic graph {$G_{X|Y}$} of $X$ given $Y$ is the undirected graph where the vertex set is $\mathcal{X}$ and the edge set is defined as follows: Two vertices $x_i \in \mathcal{X}$ and $x_j \in \mathcal{X}$ are connected by an edge if there exists a $y \in \mathcal{Y}$ such that $P_{XY}(x_i,y)\cdot P_{XY}(x_j,y) > 0$ and $f(x_i,y) \neq f(x_j,y)$.
\item An independent set of a conditional characteristic graph $G_{X|Y}$ is defined as a subset of vertices where no two vertices are connected by an edge. Let $\Gamma(G_{X|Y})$ represent the collection of independent sets. Moreover, a maximal independent set is defined as an independent set that is not a subset of any other independent set, and   $\Gamma^*(G_{X|Y})$ denotes the collection of maximal independent sets of  $G_{X|Y}$.
\item The conditional graph entropy is defined as 
\begin{align*}
    H_{G_{X|Y}}(X|Y)\! &\triangleq\! \min_{\substack{U-X-Y\\X \in U \in  \Gamma(G_{X|Y})}} I(X;U|Y)\!=\! \min_{\substack{U-X-Y\\X \in U \in  \Gamma^*(G_{X|Y})}} I(X;U|Y),
\end{align*}
where following \cite{orlitsky-2001}, we use the notation $X \in U$ when $\mathbb{P}[X \in U] = 1$ for  two random variables $X$ and $U$, where $X$ takes values in $\mathcal{X}$ and $U$ ranges over subsets of $\mathcal{X}$.
\end{enumerate}
\end{Definition}

\begin{Lemma} \label{lemma-1111} Let $\Tilde{\Omega}'_o$ be the set of probability distributions $P_{VXX_sY}$ over $\mathcal{V}\times \X \times \X_s\times \Y$ such that $P_{VXX_sY} = P_{XX_sY}P_{V|X}$ and $X \in V \in \Gamma(G_{X|Y})$, where $|\mathcal{V}| \le |\mathcal{X}| + 2$. Define
\begin{align*}
\mathcal{R}'_o \triangleq \bigcup_{P_{VXX_sY} \in \Tilde{\Omega}'_o}\Big\{(R,L) \in \mathbb{R}^2_+: 
    R \ge I(X;V|Y),~~
    L \ge I(X_s;V|Y)\Big\}.
\end{align*}
It holds that
\begin{align*}
    \mathcal{R}'_o=\mathcal{R}_o.
\end{align*}
\end{Lemma}
\begin{proof}
See Appendix \ref{appx-aa}.
\end{proof}

We are now ready to present the examples.

\begin{Example} \label{ex-1111}
By Lemma \ref{lemma-1111}, when there are no private attributes, i.e., $\S = \emptyset$, and $\P$  is a singleton, Theorem~\ref{th1} is equivalent to \cite[Th.~1]{orlitsky-2001}, where the optimal communication rate is expressed as $H_{G_{X|Y}}(X|Y)$.
\end{Example}

\begin{figure}[!t]
\centering
   \includegraphics[scale=0.6]{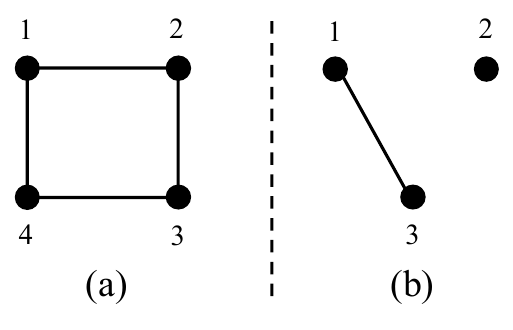}
   \caption{{Figures (a) and (b) are visual representations of the conditional characteristic graphs ${G_{X|Y}}$ for Examples \ref{ex1} and \ref{ex2}, respectively}.}
   \label{ind-set}
   \vspace{-4mm}
\end{figure}

\begin{Example} \label{ex1} Let $\mathcal{X}=\mathcal{X}_s=\{1,2,3,4\}$ and $\mathcal{Y}=\{1,2,3\}$. Assume that $X$ is uniformly distributed over the set $\mathcal{X}$ and the joint distribution in \eqref{joint-dist} is decomposed as $P_{XX_sY} = P_XP_{X_s|X}P_{Y|X_s}$. We assign the conditional probabilities $P_{X_s|X}$ and $P_{Y|X_s}$ as
\begin{align*}
\begin{bmatrix}
0.6 & 0.2 & 0.1 & 0.1 \\
0.1 & 0.6 & 0.2 & 0.1 \\
0.1 & 0.1 & 0.6 & 0.2 \\
0.2 & 0.1 & 0.1 & 0.6
\end{bmatrix}
{\rm and}
\begin{bmatrix}
0.7 & 0.2 & 0.1 \\
0.1 & 0.7 & 0.2 \\
0.2 & 0.1 & 0.7 \\
1/3 & 1/3 & 1/3
\end{bmatrix},
\end{align*}
respectively. Define the function $f(X,Y) = (X+Y)~{\rm mod}~2$. Then, the capacity region  is
\begin{align*}
\mathcal R=\{(R,L)\in\mathbb R_+^2:
R\ge H(U\mid Y),\quad L\ge I(X_s;U\mid Y)\},
\end{align*}
where $ H(U|Y) \approx 0.9820$ {bits/symbol}, and $I(X_s;U|Y) \approx 0.1007$ {bits/symbol}.
\end{Example}
\begin{proof}
From 1) and 2) of Definition \ref{def-3}, as shown in Fig. \ref{ind-set}(a), we can determine the set $\Gamma(G_{X|Y})=\{{\emptyset},\{1\},\{2\},\{3\},\{4\},\{1,3\},\{2,4\}\}$, and thus $\Gamma^*(G_{X|Y})=\{\{1,3\},\{2,4\}\}$. Let us denote $u_1 = \{1,3\}$ and $u_2 = \{2,4\}$ and $P_{U|X}(u_i|x)$ for $i \in \{1,2\}$ and $x \in \mathcal{X}$ denotes the transition probability from $X=x$ to $U=u_i$ such that $\sum_{u_i \owns x} P_{U|X}(u_i|x) = 1$. Hence, we have
\begin{align*}
P_{U|X}(u_1|1) = P_{U|X}(u_1|3) = 1~~{\rm and}~~P_{U|X}(u_1|2) = P_{U|X}(u_1|4) = 0, \\
P_{U|X}(u_2|2) = P_{U|X}(u_2|4) = 1~~{\rm and}~~P_{U|X}(u_2|1) = P_{U|X}(u_2|3) = 0.
\end{align*}
Then, using the values of the conditional distribution $P_{U|X}$ above, one can calculate
\begin{align*}
    I(X;U|Y) &\overset{\rm (a)}= H(U|Y) - H(U|X) \nonumber \displaybreak[0]\\
    &\approx 0.9820~\text{[bits/symbol]},  \nonumber \displaybreak[0]\\
    I(X_s;U|Y) &= H(U|Y) - H(U|X_s,Y) \nonumber \displaybreak[0]\\
    &\overset{\rm (b)}= H(U|Y) - H(U|X_s) \nonumber \displaybreak[0]\\
    &\approx 0.1007~\text{[bits/symbol]}.
\end{align*}
where (a) and (b) {are due to the fact that $U$ depends only on $X$ and thus the joint distribution becomes $P_{UXX_sY} = P_{U|X}P_XP_{X_s|X}P_{Y|X_s}$, which implies the Markov chain $U-X_s-Y$}.
\end{proof}

\begin{Example} \label{ex2}
Suppose that $X,X_s,Y \in \{1,2,3\}$. Define
\begin{align*}
P_{XX_sY}(x,x_s,y) &= \begin{cases}
0,~{\rm if}~ x = ~y~\\
 1/18,~{\rm if}~x~\neq~y
\end{cases}, \\
f(x,y) &= \begin{cases}
1,~{\rm if}~ x > y\\
0,~{\rm if}~ x < y
\end{cases}.
\end{align*}
Then, the capacity region is
\begin{align*}
\mathcal R=\left\{(R,L)\in\mathbb R_+^2:
R\ge\frac56-\frac12\log_2\frac32,\quad L\ge0\right\}.
\end{align*}
\end{Example}
\begin{proof}
As shown in Fig. \ref{ind-set}(b), using 1) and 2) in Definition \ref{def-3}, we can verify that only the nodes $1$ and $3$ are connected by an edge. Therefore, the set of independent sets is determined as $\Gamma({G_{X|Y}})=\{{\emptyset},\{1\},\{2\},\{3\},\{1,2\},\{2,3\}\},$ and the set of maximal independent sets is $\Gamma^*({G_{X|Y}})=\{\{1,2\},\{2,3\}\}$. Like the previous example, we denote $u_1 = \{1,2\}$ and $u_2 = \{2,3\}$ for brevity. Based on 3) in Definition \ref{def-3}, it suffices to consider the minimization of $U$ over the set $\Gamma^*({G_{X|Y}})$. Without loss of generality, one can assign the transition probabilities
\begin{align*}
P_{U|X}(u_1|1) &= P_{U|X}(u_2|3) = 1, P_{U|X}(u_1|3) = P_{U|X}(u_2|1) = 0\nonumber \\
P_{U|X}(u_1|2) &= \gamma~~{\rm and}~~P_{U|X}(u_2|2) = 1-\gamma
\end{align*}
for some $\gamma \in [0,1]$.

One can show that
$H(U|X) = \frac{1}{3}H_b(\gamma)$ and $H(U|Y) = H(U|X_s,Y) = \frac{1}{6}H_b(\gamma) - \frac{1}{6}g(\gamma) + 1$, where $g(\gamma) = (1+\gamma)\log(1+\gamma) + (2-\gamma)\log(2-\gamma)$ and $H_b(\cdot)$ represents the binary entropy function \cite{GK}. Therefore, $I(X;U|Y) = H(U|Y) - H(U|X) = 1 - \frac{1}{6}(H_b(\gamma) + g(\gamma))$, and $I(X_s;U|Y) = H(U|Y) - H(U|X_s,Y) = 0$. Since the sum $H_b(\gamma)+g(\gamma)$ is maximized at $\gamma = 0.5$, we have $I(X;U|Y) \ge \frac{5}{6} - \frac{1}{2}\log(3/2) \approx 0.5408$ [bits/symbol].
\end{proof}

\subsection{Lossy Function Computation} \label{sect-3b}

First, we present the capacity region for general discrete sources.
\begin{Theorem}[Capacity region] \label{th2} Let $\Tilde{\Omega}_D$ be the set of probability distributions $P_{UX_{\P}{X_{\S}}Y}$ over $\mathcal{U}\times \X_{\P}\times \X_{\S}\times \Y$ such that $P_{UX_{\P}X_{\S}Y} = P_{X_{\P}X_{\S}Y}P_{U|X_{\P}}$, where  $|\mathcal{U}| \le |\mathcal{X}_{\P}| + 2$.  The capacity region is
\begin{align}
\mathcal{R}_D = \bigcup_{P_{UX_{\P}X_{\S}Y} \in \Tilde{\Omega}_D}&\Big\{(R,L,D) \in \mathbb{R}^3_+:
    R \ge I(X_{\P};U|Y), \nonumber \\
    L &\ge I(X_{\S};U|Y),~~
    D \ge \mathbb{E}[d(f(X_{\P},Y),g(U,Y))] \nonumber \\
    &~~~~\text{for~some~reconstruction~function}~ {g(\cdot,\cdot)}\Big\}. \label{th2-ex}
\end{align}

\end{Theorem}
\begin{proof}
The achievability is obtained by setting $U_\I = U_{\A} = U$ and $U_{\A^c} = Q = \emptyset$ in Theorem \ref{inner-lemma}. The converse follows similarly to the proof of Theorem \ref{outer-lemma} by defining $U_t = (M, X^{t-1}_{\S}, Y^{n \backslash t})$ for $t \in [1:n]$. The cardinality bound follows by the support lemma \cite[Appx.~C]{GK}.
\end{proof}

\begin{Example}
When the target function is defined as $f(X_{\P},Y) \triangleq X_{\P}$ and an equivocation constraint $H(X^n_{\S}|M,Y^n) \ge n(E - \delta)$ is imposed in place of \eqref{pri-def2}, the leakage rate in \eqref{th2-ex} is then replaced by $E \le H(X_{\S}|U,Y)$ and as a result, Theorem~\ref{th2} recovers \cite[Th. 2 and Remark 5]{Sankar-2013} in which the encoder observes only the public attributes.
\end{Example}

\begin{Example}
In a special case, where $\S = \emptyset$ and $|\P| = 1$, Theorem \ref{th2} reduces to \cite[Th.~1]{Yamamoto-fc}.
\end{Example}

Next, we apply the single-letter characterization obtained in Theorem~\ref{th2} to investigate the capacity region of Gaussian sources for two different setups. Although Theorem~\ref{th2} is derived for discrete sources, it can also be extended to Gaussian sources by employing a fine quantizer, e.g.,  \cite{rana2022,chou2014separation}. As in Examples \ref{ex1} and \ref{ex2}, let both $\P$ and $\S$ be singletons. Our first setup adopts the same joint distribution as in \cite[Ex.~2]{Sankar-2013}.

\begin{Setup} \label{setup-1} Consider $X \sim \mathcal{N}(0,\sigma^2_{X})$ {with $\sigma^2_{X} > 0$}, and the conditional distributions $P_{X_s|X}$ and $P_{Y|X}$ are modeled as
\begin{align}
X_s \triangleq X + N_{X_s},~~~Y \triangleq X + N_Y, \label{setup2-relation}
\end{align}
where $N_{X_s} \sim \mathcal{N}(0,\sigma^2_{N_{X_s}})$ and $N_Y \sim \mathcal{N}(0,\sigma^2_{N_Y})$ {with $\sigma^2_{N_{X_s}} > 0 $ and $\sigma^2_{N_Y} > 0$. $N_{X_s}$ and $N_Y$ are independent of $X$ and each other}. In addition, the distortion criterion is quantified by the quadratic distortion measure, i.e., $d(f,\hat{f}) = (f-\hat{f})^2$, and the target function is computed as $F \triangleq a(X + Y)$ with $0 < a < \infty$. For instance, the defined function gives the mean between the public attribute and side information if the constant $a$ is $\frac{1}{2}$.
\end{Setup} 

\begin{Theorem}[Capacity region] \label{th3}
For a given $D > 0$, the capacity region for Setup \ref{setup-1} is
\begin{align}
\mathcal{R}_D(D) = \Big\{(R,L) \in \mathbb{R}^2_+:
    R &\ge \frac{1}{2}\log^{+} \Big(\frac{a^2(1-\rho^2_{XY})\sigma^2_X}{D}\Big), \nonumber \\
    L &\ge \frac{1}{2}\log^{+}\Big(\frac{1-\rho^2_{XX_s}\rho^2_{XY}}{1-\rho^2_{XX_s} + \rho^2_{XX_s}D/(a^2\sigma^2_{X})}\Big)\Big\}, \label{th3-eq}
\end{align}
where {$\log^{+}(x) = \max\{0,\log(x)\}$ for $x \in \mathbb{R}_+$}, $\rho^2_{XX_s} = \sigma^2_X/(\sigma^2_X + \sigma^2_{N_{X_s}})$, and $\rho^2_{XY} = \sigma^2_X/(\sigma^2_X + \sigma^2_{N_Y})$.
\end{Theorem}
\begin{proof}
The achievable region is shown by choosing $U$ to be a Gaussian random variable. For the converse, we make use of the maximum entropy principle along with the monotonicity of the logarithmic function. See Appendix~\ref{appx-bb} for the complete proof.
\end{proof}

Special cases of Theorem \ref{th3} are outlined below.

\begin{Example}
Assume
$0<D\le a^2(1-\rho_{XY}^2)\sigma_X^2$. If the leakage metric \eqref{pri-def2} is replaced by the unconditional mutual information $I(X^n_s;M,Y^n) \le n(L+\delta)$ as in \cite[Ex. 2]{Sankar-2013}, the leakage rate becomes $L \ge I(X_s;U,Y) = \frac{1}{2}\log^+\frac{1}{1-\rho^2_{XX_s} + \rho^2_{XX_s}D/(a^2\sigma^2_{X})}$. Consequently, in the special case where $a =1$ and the unconditional mutual information is used as the leakage metric, Theorem \ref{th3} matches the capacity region of \cite[Case (ii) in Ex.~2]{Sankar-2013}.
\end{Example}

\begin{Example}
When $a=1$ and the decoder has no side information ($\rho_{XY} = 0$), Theorem~\ref{th3} coincides with the capacity region derived in \cite[Case (i) in Ex.~2]{Sankar-2013}.
\end{Example}

\begin{Example} When $a = 1$ and there is no leakage constraint, Theorem \ref{th3} reduces to \cite[Remark~6]{wyner-ziv}, where for a given distortion $D > 0$, we have $R \ge \frac{1}{2}\log^+\frac{(1 - \rho^2_{XY})\sigma^2_{X}}{D} = \frac{1}{2}\log^+\frac{\sigma^2_{X}\sigma^2_{N_Y}}{(\sigma^2_{X} + \sigma^2_{N_Y})D}$.
\end{Example}

In \eqref{setup2-relation}, both the private attribute $X_s$ and side information $Y$ were modeled as noisy versions of the public attribute $X$. We next explore a different setup (Setup \ref{setup-2}), where $X$ and $Y$ serve instead as noisy observations of $X_s$. This model reflects the hidden-private-data scenario \cite[Fig.~1]{amirraza-2024}, where the agent (encoder) only observes noisy measurements rather than sensitive data. Due to the different correlation among $(X,X_s,Y)$ in Setups \ref{setup-1} and \ref{setup-2}, the technique used to derive the capacity region for Setup \ref{setup-1} does not seem easily applicable to Setup \ref{setup-2}, especially the analysis of the leakage rate.

\begin{Setup} \label{setup-2} The target function and distortion measure remain the same as in Setup \ref{setup-1}. However, in this setup, we consider
\begin{align}
X = X_s + \Tilde{N}_X,~~~Y = X_s + \Tilde{N}_Y, \label{xxsy-eq}
\end{align}
where $X_s \sim \mathcal{N}(0,\sigma^2_{X_s})$, $\Tilde{N}_X \sim \mathcal{N}(0,\sigma^2_{\Tilde{N}_X})$, and $\Tilde{N}_Y \sim \mathcal{N}(0,\sigma^2_{\Tilde{N}_Y})$ with $\sigma^2_{X_s} > 0$, $\sigma^2_{\Tilde{N}_X} > 0$, and $\sigma^2_{\Tilde{N}_Y} > 0$. $\Tilde{N}_X$ and $\Tilde{N}_Y$ are independent of $X_s$ and each other.
\end{Setup}

\begin{figure*}[t]
  \centering
    \includegraphics[width=\linewidth]{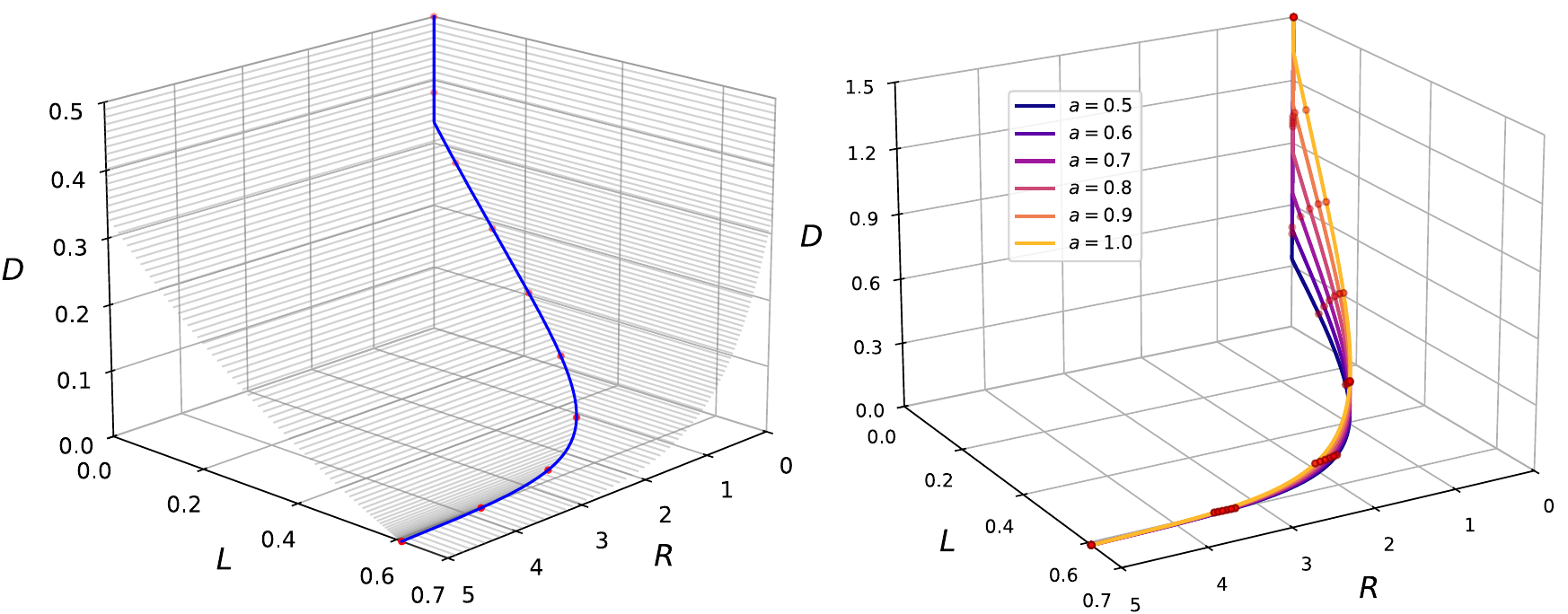}
    \vspace{-10mm}
  \caption{(a) illustrates the rate region of tuple $(R,L,D)$ in Theorem \ref{th3} when the source obeys a Gaussian distribution and the computed function is defined as $F = \frac{1}{2}(X+Y)$. The achievable region is the upper part of the region bounded by the light-gray surfaces. (b) depicts the curves of $(R,L,D)$ with different target functions by varying the parameter $a$.}
  \label{aa}
\end{figure*}

\begin{Theorem}[Capacity region] \label{th3-222}
The capacity region for Setup \ref{setup-2} is
\begin{align}
\mathcal{R}_D = \bigcup_{0 < \alpha \le 1 }\Big\{(R,L,D) \in \mathbb{R}^3_+:
R &\ge \frac{1}{2}\log\Big(\frac{\alpha\rho^2_{XY} + 1-\rho^2_{XY}}{\alpha}\Big), \nonumber \displaybreak[0] \\
L &\ge \frac{1}{2}\log\Big(\frac{\alpha\rho^2_{XY} + 1-\rho^2_{XY}}{\alpha\rho^2_{XX_s} + 1 -\rho^2_{XX_s}}\Big), \nonumber \displaybreak[0] \\
D &\ge \frac{\alpha a^2(1-\rho^2_{XY})(\sigma^2_{X_s} + \sigma^2_{\Tilde{N}_X})}{\alpha\rho^2_{XY} + 1-\rho^2_{XY}}
\Big\},
\label{sbd-1}
\end{align}
where $\rho^2_{XY} = \rho^2_{XX_s}\rho^2_{X_sY}$ with $\rho^2_{XX_s} = \sigma^2_{X_s}/(\sigma^2_{X_s} + \sigma^2_{\Tilde{N}_X})$ and $\rho^2_{X_sY} = \sigma^2_{X_s}/(\sigma^2_{X_s} + \sigma^2_{\Tilde{N}_Y})$.
\end{Theorem}
\begin{proof}
The achievability follows by choosing $U$ as a Gaussian random variable, while the converse is shown via a Fisher information analysis, e.g., \cite{ekrem-2014}. The detailed proof is provided in Appendix~\ref{appx-bb222}.
\end{proof}

To illustrate Theorem \ref{th3}, we present a numerical example using $\sigma^2_X = 4$, $\sigma^2_{N_{X_s}} = 1$, and $\sigma^2_{N_Y} = 2$. In Fig. \ref{aa}(a), we fix $a = 1/2$ and the blue curve traces the optimal points defined by the communication and leakage rates in \eqref{th3-eq} across various distortion levels. As distortion approaches zero, the required communication rate is large, and,  conversely, higher distortion requires less communication rate. This behavior highlights the trade-off: achieving smaller distortion requires downloading more information from the database, which in turn increases the correlation between the response and the private attribute, contributing to privacy leakage. The sharp vertical boundary appears because communication becomes unnecessary $R = 0$ (and $L = 0$) once the distortion exceeds $a^2(1-\rho^2_{XY})\sigma^2_X = 1/3$. Finally, Fig. \ref{aa}(b) illustrates the $(R,L,D)$ tuples for different target functions, obtained by varying values of the parameter $a$. The graphs show that larger values of $a$ yield higher distortion under the same communication rate $R$. This is because increasing $a$ raises the variance of the target function, inherently leading to higher estimation error at the same communication cost.

\section{Distributed Databases} \label{sect4}
In this section, we investigate the capacity region of a system where all private and public attributes are distributively stored in $\kappa~(\ge 2)$ different DBs.
For any nonempty set $\A \subseteq \I$, we denote {$\P_{\A} \triangleq \bigcup_{i \in \A}\P_i$, $\S_{\A} \triangleq \bigcup_{i \in \A}\S_i$}, and $R_{\A} \triangleq \sum_{i\in\A}R_i$.

\subsection{Distributed Lossless Function Computation}

For lossless settings, we have the following results.

\begin{Theorem}[Inner Bound] \label{inner-bound} Let $\Omega_{\rm in}$ be the set of probability distributions $P_{U_{\I}X_{\P}X_{\S}YQ}$ over $\mathcal{U}_{\I}\times {\X_{\P}}\times \X_{\S}\times \Y \times \mathcal{Q}$ such that
\begin{align}
P_{U_{\I}X_{\P}X_{\S}YQ} = P_{Q}P_{X_{\P}}P_{X_{\S}Y|X_{\P}}\prod_{i=1}^{\kappa}P_{U_i|X_{\Pi}Q} \label{inner-mk}
\end{align}
and $H(F|U_{\I},Y,Q) = 0$, where the time-sharing random variable $Q$ is independent of the source variables $(X_\P,X_\S,Y)$. An inner bound on the capacity region $\mathcal{R}$ is given by
\begin{align}
\mathcal{R} \supseteq \bigcup_{P_{U_{\I}X_{\P}X_{\S}YQ} \in \Omega_{\rm in}}&\Big\{
((R_i)_{i \in \I},L) \in \mathbb{R}^{\kappa + 1}_+: \nonumber \\
R_{\A} &\ge I({X_{\P_\A}};U_{\A}|U_{\A^c},Y,Q),~~~\forall \A \subseteq \I,  \nonumber \\
L &\ge I(X_{\S};U_{\I}|Y,Q)
\Big\}. \label{inner-ll}
\end{align}
\end{Theorem}
\begin{proof}
The proof consists of two steps. First, we show that it suffices to establish the achievability of a corner point on the dominant face of the region defined in Theorem \ref{inner-bound} for its achievability. We then demonstrate the achievability of this corner point using the output statistics of random binning (OSRB) technique \cite{Yassaee-tit2014}. See Appendix~\ref{app-b-a} for details.
\end{proof}

\begin{Theorem}[Outer Bound] \label{outer-bound} Let $\Omega_{\rm out}$ be the set of probability distributions $P_{U_{\I}X_{\P}X_{\S}YQ}$ over $\mathcal{U}_{\I}\times \X_{\P}\times \X_{\S}\times \Y \times \mathcal{Q}$ such that
\begin{align}
(Q,U_i) - X_{\P_i}-(X_{\P\setminus \P_i},X_\S,Y),~\forall i \in \I, \label{outer-mk}
\end{align}
and $H(F|U_{\I},Y,Q) = 0$, {where $Q$ is independent of $(X_\P,X_\S,Y)$}. An outer bound on the capacity region $\mathcal{R}$ is
\begin{align}
\mathcal{R} \subseteq \bigcup_{P_{U_{\I}X_{\P}X_{\S}YQ} \in \Omega_{\rm out}}&\Big\{((R_i)_{i \in \I},L) \in \mathbb{R}^{\kappa + 1}_+: \nonumber \\
R_{\A} &\ge I(X_{\P_\A};U_{\A}|U_{\A^c},Y,Q) \nonumber \\
&~~~~+[I(U_{\A};U_{\A^c}|Y,Q)- I(U_{\A};U_{\A^c}|X_{\P_\A},Y,Q)]^-,~~{\forall \A \subseteq \I}, \nonumber \\
L &\ge I(X_{\S};U_{\I}|Y,Q)
\Big\}, \label{outer-ll}
\end{align}
where $[x]^- = \min\{0,x\}$ for $x \in \mathbb{R}$.
\end{Theorem}
\begin{proof}
    The proof follows from standard converse techniques involving Fano's inequality and the introduction of auxiliary random variables. See Appendix~\ref{app-b-b} for details.
\end{proof}

The inner bound in Theorem~\ref{inner-bound} and the outer bound in Theorem~\ref{outer-bound} do not match due to the difference between the Markov chains defined in \eqref{inner-mk} and \eqref{outer-mk}. As a result of this, the communication rates are bounded differently. In particular, since the conditions of the Markov chains defined by \eqref{outer-mk} are less restrictive, the outer bound optimizes over a larger family of joint distributions $P_{U_{\mathcal{I}}X_{\mathcal{P}}X_{\mathcal{S}}YQ}$ than the inner bound.
When the databases and side information are mutually independent, a separate converse
exploiting this independence matches the inner bound and yields the capacity region stated
below.

\begin{Theorem}[Capacity region] \label{th3-3-ll} Let $\Omega$ be the set of probability distributions $P_{U_{\I}X_{\P}X_{\S}YQ}$ over $\mathcal{U}_{\I}\times \X_{\P}\times\X_{\S}\times \Y\times\mathcal{Q}$ such that
\begin{align}
P_{U_{\I}X_{\P}X_{\S}YQ} = {P_Q}P_Y\prod_{i=1}^{\kappa}P_{X_{\Pi}X_{\Si}}P_{U_i|X_{\Pi}Q} \label{th6-mk}
\end{align}
and $H(F|U_{\I},Y,Q) = 0$, {where $Q$ is independent of $(X_\P,X_\S,Y)$}. The capacity region is  
\begin{align}
\mathcal{R} = \bigcup_{P_{U_{\I}X_{\P}X_{\S}YQ} \in \Omega}\Big\{((R_i)_{i \in \I},L) \in \mathbb{R}^{\kappa + 1}_+: R_i &\ge I(X_{\Pi};U_i|Q), \forall i \in \I, \nonumber \\
L &\ge I(X_{\S};U_{\I}|Q)\Big\}. 
\label{indep-eq-ll}
\end{align}
\end{Theorem}
\begin{proof}
The achievability and converse proofs follow along the same lines as those of Theorems \ref{inner-bound} and \ref{outer-bound}, respectively. See Appendix~\ref{app-c} for details.
\end{proof}

\begin{Lemma}[Cardinality bounds] \label{CB-ll} It holds that
\begin{enumerate}
    \item[{\rm (a)}] In Theorem \ref{inner-bound}, the cardinality bound of $U_i$ can be limited to $|\mathcal{U}_i| \le |\mathcal{X}_{\Pi}| + 2^{\kappa + 1} - 1$ for $i \in \I$.
    \item[{\rm (b)}] {In Theorem \ref{outer-bound}, it suffices to limit $|\mathcal{U}_i| \le |\mathcal{F}|$ for $i \in \I$}.
    \item[{\rm (c)}] In Theorem \ref{th3-3-ll}, $|\mathcal{U}_i| \le |\mathcal{X}_{\Pi}| + 2$ for $i \in \I$.
    \item[{\rm (d)}] In Theorems~\ref{inner-bound}, \ref{outer-bound}, and \ref{th3-3-ll}, the cardinality bound of $Q$ satisfies $|\mathcal{Q}| \le \kappa + 2$.
\end{enumerate}
\end{Lemma}
\begin{proof}
{We apply the support lemma \cite[Appx.~C]{GK} to prove (a) and (c), and the Fenchel–Eggleston–Carathéodory theorem \cite[Appx.~A]{GK} for (d). For (b), we provide a novel proof based on replacing auxiliary variables. Complete proofs are provided in Appendix \ref{Appx-CB-ll}}.
\end{proof}

\subsection{Distributed Lossy Function Computation}

Inner and outer bounds on the capacity region $\mathcal{R}_D$ for general discrete sources are presented below.
\begin{Theorem}[Inner bound] \label{inner-lemma} Let $\bar{\Omega}_{\rm in}$ be the set of probability distributions $P_{U_{\I}X_{\P}X_{\S}YQ}$ over $\mathcal{U}_{\I}\times \X_{\P}\times\X_{\S}\times \Y \times \mathcal{Q}$ such that
\begin{align}
P_{U_{\I}X_{\P}X_{\S}YQ} \triangleq \textstyle P_{Q}P_{X_{\P}}P_{YX_{\S}|X_{\P}}\prod_{i=1}^{\kappa}P_{U_i|X_{\Pi}Q}, \label{inner-joint-ls}
\end{align}
where $Q$ is independent of $(X_\P,X_\S,Y)$. An inner bound is given by
\begin{align}
\mathcal{R}_D \supseteq \bigcup_{P_{U_{\I}X_{\P}X_{\S}YQ} \in \bar{\Omega}_{\rm in}}&\Big\{((R_i)_{i \in \I},L,D) \in \mathbb{R}^{\kappa + 2}_+: \nonumber\\ R_{\A} &\ge I(X_{\P_\A};U_{\A}|U_{\A^c},Y,Q),~~\forall \A \subseteq \I, \nonumber \\
L &\ge I(X_{\S};U_{\I}|Y,Q), \nonumber \\
D & \ge \mathbb{E}[d(f(X_{\P},Y),g(U_{\I},Y,Q))]\nonumber \\
&\text{for some reconstruction function}~{g(\cdot,\cdot,\cdot)}\Big\}. \label{ls-inner-eq}
\end{align}
\end{Theorem}
\begin{proof}
The proof employs the same coding scheme as in Theorem \ref{inner-bound} and follows similarly to that of Theorem \ref{inner-bound} except for the additional analysis of the distortion constraints. We therefore provide only this additional analysis. See Appendix~\ref{appx-cc-1} for details.
\end{proof}

\begin{Theorem}[Outer bound] \label{outer-lemma} Let $\bar{\Omega}_{\rm out}$ be the set of probability distributions $P_{U_{\I}X_{\P}X_{\S}YQ}$ over $\mathcal{U}_{\I}\times \X_{\P}\times\X_{\S}\times \Y \times \mathcal{Q}$ such that
\begin{align}
    (Q,U_i)-X_{\Pi}-(X_{\P \backslash \Pi},X_{\S},Y),~\forall i \in \I, \label{mk-ls}
\end{align}
where $Q$ is independent of $(X_\P,X_\S,Y)$. An outer bound is
\begin{align}
\mathcal{R}_D \subseteq \bigcup_{P_{U_{\I}X_{\S}X_{\P}YQ} \in \bar{\Omega}_{\rm out}}&\Big\{((R_i)_{i \in \I},L,D) \in \mathbb{R}^{\kappa + 2}_+: \nonumber\\
R_{\A} &\ge {I(X_{\P_\A};U_{\A}|U_{\A^c},Y,Q)} \nonumber \\
&~~~~{+[I(U_{\A};U_{\A^c}|Y,Q)- I(U_{\A};U_{\A^c}|X_{\P_\A},Y,Q)]^-},~~{\forall \A \subseteq \I}, \nonumber \\
L &\ge I(X_{\S};U_{\I}|Y,Q), \nonumber \\
D & \ge \mathbb{E}[d(f(X_{\P},Y),g(U_{\I},Y,Q))]\nonumber \\
&\text{for some reconstruction function}~g(\cdot,\cdot,\cdot)\Big\}.
\label{ls-th-eq}
\end{align}
\end{Theorem}
\begin{proof}
We give only the analysis of the distortion constraint as the other constraints follow similarly to those in Theorem \ref{outer-bound}. See Appendix~\ref{appx-cc-2} for details.
\end{proof}

Parallel to the distributed lossless scenario, the mismatch between Theorems~\ref{inner-lemma} and \ref{outer-lemma} stems from the distinction between the Markov chains defined in \eqref{inner-joint-ls} and \eqref{mk-ls}. However, for mutually independent DBs and side information, we obtain the capacity region stated below.

\begin{Theorem}[Capacity region] \label{ls-capacity} Let $\bar{\Omega}$ be the set of probability distributions $P_{U_{\I}X_{\P}X_{\S}YQ}$ over $\mathcal{U}_{\I}\times \X_{\P}\times\X_{\S}\times \Y\times\mathcal{Q}$ such that
$P_{U_{\I}X_{\P}X_{\S}YQ} = {P_Q}P_Y\prod_{i=1}^{\kappa}P_{X_{\Pi}X_{\Si}}P_{U_i|X_{\Pi}Q}$, where $Q$ is independent of $(X_\P,X_\S,Y)$. Then, the capacity region is
\begin{align}
\mathcal{R}_D = \bigcup_{P_{U_{\I}X_{\P}X_{\S}YQ} \in \bar{\Omega}}&\Big\{((R_i)_{i \in \I},L,D) \in \mathbb{R}^{\kappa + 2}_+: \nonumber\\
R_i &\ge I(X_{\Pi};U_i|Q), \forall i \in \I, \nonumber \\
L &\ge I(X_{\S};U_{\I}|Q), \nonumber \\
D &\ge \mathbb{E}[d(f(X_{\P},Y),g(U_{\I},Y,Q))] \nonumber \\
&\text{for some reconstruction function}~{g(\cdot,\cdot,\cdot)}\Big\}.
\label{indep-eq-ls}
\end{align}
\end{Theorem}
\begin{proof}
We omit the detailed proof of Theorem \ref{ls-capacity} as it closely follows the proof of Theorem \ref{th3-3-ll} with the analysis of the distortion constraint. Specifically, the distortion constraint in \eqref{indep-eq-ls} is analyzed using the same steps as in \eqref{d-eqls} for the achievability and \eqref{d-ana} for the converse with $U_{i,t} \triangleq (M_i, X_{\mathcal{S}_i}^{t-1})$ and $Q_t \triangleq Y^{n \setminus t}$ for $i \in \mathcal{I}$ and $t \in [1:n]$.
\end{proof}

\begin{Lemma}[Cardinality bounds] \label{CB-ls} It holds that
\begin{enumerate}
    \item[{\rm (a)}] In Theorem \ref{inner-lemma}, it suffices to limit $|\mathcal{U}_i| \le |\mathcal{X}_{\Pi}| + 2^{\kappa + 1} - 1$ for $i \in \I$.
    \item[{\rm (b)}] {In Theorem \ref{outer-lemma}, it suffices to limit $|\mathcal{U}_i| \le |\mathcal{F}|$ for $i \in \I$}.
    \item[{\rm (c)}] In Theorem \ref{ls-capacity}, it suffices to limit $|\mathcal{U}_i| \le |\mathcal{X}_{\Pi}| + 2$ for $i \in \I$.
    \item[{\rm (d)}] In Theorems~\ref{inner-lemma}, \ref{outer-lemma}, and \ref{ls-capacity}, it suffices to limit $|\mathcal{Q}| \le \kappa + 3$.
\end{enumerate}
\end{Lemma}
\begin{proof}
See Appendix \ref{Appx-CB-ls}.
\end{proof}

\section{Concluding Remarks} \label{sect5}
In this paper, we first analyzed a setup where all private and public attributes are stored in a single DB and characterized the capacity region of communication and leakage rates for lossless function computation, as well as the capacity region of communication rate, leakage rate, and distortion in lossy function computation for discrete and Gaussian sources. Subsequently, we examined a scenario where private and public attributes are distributively stored across multiple DBs, deriving inner and outer bounds on the capacity region, and characterized the capacity region when {the DBs and side information are mutually independent}. Future directions include extending our distributed database model to account for an eavesdropper, analogous to the single-database setting in \cite{va-remi-isit2026}, and investigating the capacity region for other target function structures under Gaussian sources.

\appendices

\section{Proof of Lemma \ref{lemma-1111}} \label{appx-aa}
A similar approach for proving \cite[Th.~2]{orlitsky-2001} is taken, but it is adapted to account for the leakage constraint.

First, we prove that $\mathcal{R}'_o \subseteq \mathcal{R}_o$.
Let $v \in \Gamma(G_{X|Y})$ and $y \in \Y$. We define $g(v,y) = f(x',y)$ for any $x' \in v$ such that $P_{XY}(x',y) > 0$, if such an $x'$ exists, and assign $g(v,y)$ an arbitrary value in $\mathcal{F}$ otherwise. By 1) of Definition~\ref{def-3}, it follows that
$  f(x,y) = f(x',y) = g(v,y)$ 
for all $x \in v$ such that $P_{XY}(x,y) > 0$. Consequently, $X \in V \in \Gamma(G_{X|Y})$ implies the existence of a deterministic function $g$ such that $F = g(V,Y)$ almost surely, which leads to $H(F|V,Y) = 0$.

Next, we prove the reverse inclusion $\mathcal{R}'_o \supseteq \mathcal{R}_o$. Let $U$ be a random variable satisfying the conditions of $\mathcal{R}_o$ in~\eqref{ex1-eq111}.
For $u \in \mathcal{U}$, define $v(u) = \{x: P_{UX}(u,x) > 0\}$, and construct a random variable $V$ according to
\begin{align}
    P_{V|UXX_sY}(v|u,x,x_s,y) = \begin{cases}
      1 & \text{ if}~~v = v(u) \\
      0 & \text{ otherwise}.
   \end{cases} \label{mk-appxa}
\end{align}
This construction induces the Markov chain $V-U-(X,X_s,Y)$.

We begin by establishing the Markov chain $V-X-(X_s,Y)$. Observe that
\begin{align}
    I(V;X_s,Y|X) &\le I(V,U;X_s,Y|X) \nonumber \displaybreak[0]\\
    &\overset{\rm (a)}= I(U;X_s,Y|X) \nonumber \displaybreak[0]\\
    &\overset{\rm (b)}= 0, \label{nonzero-mt}
\end{align}
where (a) follows since $V$ is a function of $U$ by \eqref{mk-appxa}, and (b) holds by the Markov chain $U-X-(X_s,Y)$ since $U$ satisfies the conditions of $\mathcal{R}_o$ in \eqref{ex1-eq111}. By the non-negativity of mutual information, \eqref{nonzero-mt} implies $I(V;X_s,Y|X)=0$.

We now verify that $X \in V \in \Gamma(G_{X|Y})$. Suppose $P_{VX}(v,x) > 0$. Then, there exists a $u$ such that $v = v(u)$ and $P_{UX}(u,x) > 0$, which implies $x \in v$ by definition of $v(u)$. Moreover, for this $u$, every $x \in v$ satisfies $P_{UX}(u,x) > 0$. Together with the Markov chain $U-X-(X_s,Y)$, this gives
$P_{UXY}(u,x,y)=P_{U|X}(u|x)P_{XY}(x,y)>0$
whenever $x\in v$ and $P_{XY}(x,y)>0$.
Since $H(F|U,Y)=0$, there exists a deterministic function $g$ such that $f(x,y)=g(u,y)$ for all such pairs $(x,y)$. Consequently, for any $x_1,x_2 \in v$ and $y$ satisfying $P_{XY}(x_1,y)P_{XY}(x_2,y)>0$, we have $f(x_1,y) = g(u,y)$ and $f(x_2,y) = g(u,y)$, therefore, $f(x_1,y) = f(x_2,y)$, which confirms that $V \in \Gamma(G_{X|Y})$.

It remains to show that $I(X;V|Y) \le I(X;U|Y)$ and $I(X_s;V|Y) \le I(X_s;U|Y)$. We note that the Markov chain $V-U-(X,X_s,Y)$ leads to
\begin{align}
V-(U,Y)-X~~\text{and}~~V-(U,Y)-X_s \label{vuyx-mk}
\end{align}
due to $I(V;X|U,Y)\overset{\rm (a)}=I(V;X,Y|U) = 0$ and $I(V;X_s|U,Y)\overset{\rm (b)}=I(V;X_s,Y|U) = 0$, where (a) and (b) hold by the Markov chain $V-U-Y$. Then, we have
\begin{align*}
I(X;V|Y) &= H(X|Y) - H(X|V,Y) \nonumber \\
&\le H(X|Y) - H(X|V,U,Y) \nonumber \displaybreak[0]\\
&\overset{\rm (a)}= H(X|Y) - H(X|U,Y) \nonumber \displaybreak[0]\\
&= I(X;U|Y), \\
I(X_s;V|Y) &= H(X_s|Y) - H(X_s|V,Y) \nonumber \displaybreak[0]\\
&\le H(X_s|Y) - H(X_s|V,U,Y) \nonumber \displaybreak[0]\\
&\overset{\rm (b)}= H(X_s|Y) - H(X_s|U,Y) \nonumber \displaybreak[0]\\
&= I(X_s;U|Y),
\end{align*}
where (a) and (b) follow from \eqref{vuyx-mk}.
\qed
\section{Proof of Theorem \ref{th3}} \label{appx-bb}

\noindent{\em Achievability}: The proof follows by considering $X = U + \Theta$, where $U \sim \mathcal{N}(0,(1-\alpha)\sigma^2_X)$ and $\Theta \sim \mathcal{N}(0,\alpha\sigma^2_X)$ for $\alpha \in (0,1]$, and choosing the reconstruction function $g(\cdot,\cdot)$ as the MMSE estimator. 

\noindent{\em Converse}: For the distortion, we have
\begin{align}
    D &\geq \mathbb{E}[(F-g(U,Y))^2] \nonumber \\ &\overset{\rm (a)}\ge \E[(F-\mathbb{E}[F|U,Y])^2] \nonumber \displaybreak[0] \\
    &\overset{\rm (b)}= \MMSE(F|U,Y) = \MMSE(a(X+Y)|U,Y) \nonumber \displaybreak[0] \\
    &= a^2\MMSE(X|U,Y), \label{d-eqeq11}
\end{align}
where (a) holds for any function $g(\cdot,\cdot)$ (cf.~\cite[eq.~(8.78)]{cover}) and (b) follows with  the notation $\MMSE(A|V) = \E_{V}[\Var[A|V]]$.

We note that \eqref{d-eqeq11} also implies that $h(X |U,Y) \le \frac{1}{2}\log(2\pi e D/a^2)$. Based on this inequality, if we fix the conditional entropy $h(X|U,Y)$ and apply the conditional EPI to $X_s = X + N_{X_s}$, we obtain a lower bound on $h(X_s|U,Y)$. However, such a lower bound cannot be used to analyze the leakage rate since bounding the leakage rate requires an upper bound on $h(X_s|U,Y)$.

In this proof, we instead exploit the maximum entropy principle together with the monotonicity of the logarithmic function to derive bounds on both the communication and leakage rates.

\smallskip
\noindent{\em Analysis of Communication Rate}: It holds that
\begin{align}
    R &\ge I(X;U|Y) = h(X|Y) - h(X|U,Y) \nonumber \displaybreak[0] \\
    &\overset{\rm (a)}\ge \frac{1}{2}\log2\pi e(1 - \rho^2_{XY})\sigma^2_X - \frac{1}{2}\log2\pi e \big(\MMSE(X|U,Y)\big) \label{r45-eqeq} \displaybreak[0] \\
    &\overset{\rm (b)}\ge \frac{1}{2}\log(1 - \rho^2_{XY})\sigma^2_X - \frac{1}{2}\log\frac{D}{a^2} \nonumber \displaybreak[0] \\
    &\ge \frac{1}{2}\log\frac{a^2(1 - \rho^2_{XY})\sigma^2_{X}}{D}, \nonumber
\end{align}
where (a) follows by the maximum entropy principle \cite[Th.~8.6.5]{cover} and Jensen's inequality and (b) follows by applying \eqref{d-eqeq11} and noting that the logarithmic function is monotonically increasing.

\smallskip
\noindent{\em Analysis of Leakage Rate}: We have
\begin{align}
    L &\ge I(X_s;U|Y) = h(X_s|Y)-h(X_s|U,Y) \nonumber \displaybreak[0] \\
    &\overset{\rm (a)}\ge \frac{1}{2}\log2\pi e(1-\rho^2_{XX_s}\rho^2_{XY})(\sigma^2_X +\sigma^2_{N_{X_s}})- \frac{1}{2}\log2\pi e\big(\MMSE(X_s|U,Y)\big) \nonumber \displaybreak[0] \\
    &\overset{\rm (b)}= \frac{1}{2}\log(1-\rho^2_{XX_s}\rho^2_{XY})(\sigma^2_X +\sigma^2_{N_{X_s}})- \frac{1}{2}\log\left(\MMSE(X|U,Y) + \sigma^2_{N_{X_s}}\right) \label{4747-eq} \\
    &\overset{\rm (c)}\ge \frac{1}{2}\log(1-\rho^2_{XX_s}\rho^2_{XY})(\sigma^2_X +\sigma^2_{N_{X_s}})- \frac{1}{2}\log(\frac{D}{a^2} + \sigma^2_{N_{X_s}}) \nonumber \displaybreak[0] \\
    &= \frac{1}{2}\log\frac{1-\rho^2_{XX_s}\rho^2_{XY}}{1-\rho^2_{XX_s} + \rho^2_{XX_s}D/(a^2\sigma^2_{X})}, \nonumber
\end{align}
where (a) is due to \cite[Th.~8.6.5]{cover} and Jensen's inequality, (b) holds due to the independence between $N_{X_s}$ and $(U,Y)$ as {$N_{X_s}$ is independent of $X$ and $N_Y$}, and (c) follows from \eqref{d-eqeq11} and because the logarithm is increasing.
\qed

\section{Proof of Theorem \ref{th3-222}} \label{appx-bb222}

\noindent{\em Achievability}: The proof follows by considering $X = U + \Theta$, where $U \sim \mathcal{N}(0,(1-\alpha)(\sigma^2_{X_s} + \sigma^2_{\Tilde{N}_X}))$ and $\Theta \sim \mathcal{N}(0,\alpha(\sigma^2_{X_s} + \sigma^2_{\Tilde{N}_X}))$ for $\alpha \in (0,1]$, and choosing $g(\cdot,\cdot)$ as the MMSE estimator.

\smallskip
\noindent{\em Converse}: {Note that the auxiliary random variable $U \sim P_{U|X}$ and the source random variables $(X,X_s,Y)$ form the Markov chain
\begin{align}
U - X - X_s - Y. \label{long-mk}
\end{align}
Moreover, by \cite[Th.~3.5.2]{gallager-2013}, from \eqref{xxsy-eq}, we can express
\begin{align}
    X_s = \nu X + \tilde{N}_{X_s},~~Y
    = X_s + \Tilde{N}_Y = \nu X + N'_Y,
    \label{xseq111}
\end{align}
where $\nu = \sigma^2_{X_s}/(\sigma^2_{X_s} + \sigma^2_{\Tilde{N}_X})$, and
$\tilde{N}_{X_s} \sim \mathcal{N}(0,(1-\nu)\sigma^2_{X_s})$ and $N'_Y \sim \mathcal{N}(0,(1-\nu)\sigma^2_{X_s} + \sigma^2_{\Tilde{N}_Y})$, independent of $X$. Throughout this proof, we work with \eqref{xseq111} rather than \eqref{xxsy-eq} to derive the outer bound of \eqref{sbd-1} as it facilitates the application of Fisher-information identities}.

Before proceeding to the proof, we note that the method used for showing the converse proof of Theorem \ref{th3} cannot be directly extended to this setup because the independence structure in \eqref{4747-eq} fails to hold for \eqref{xseq111}. Specifically, $\tilde{N}_{X_s}$ and $(U,Y)$ are not independent as $\tilde{N}_{X_s}$ depends on $Y$ through $X_s$. Likewise, in \eqref{xxsy-eq}, $\tilde{N}_{X}$ depends on $U$ via $X$. Therefore, we adopt a different approach in this proof. The details are presented below.

\noindent{\em Analyses of Distortion and Communication rates}:~~~Similar to \eqref{d-eqeq11} and \eqref{r45-eqeq}, we have
\begin{align}
    D &\ge a^2\MMSE(X|U,Y), \label{d-middle-con} \\
    R &\ge \frac{1}{2}\log \frac{(1-\rho^2_{XY})(\sigma^2_{X_s} + \sigma^2_{\Tilde{N}_X})}{\MMSE(X|U,Y)}. \label{r-middle-con}
\end{align}

\noindent{\em Analysis of Leakage Rate}: It follows that
\begin{align}
    L &\ge I(X_s;U|Y) \overset{\rm (a)}= I(X_s;U) - I(Y;U) \nonumber \displaybreak[0]\\
    &= \frac{1}{2}\log\frac{\sigma^2_{X_s}}{\sigma^2_{X_s} + \sigma^2_{\Tilde{N}_Y}} + h(Y|U) - h(X_s|U), \label{l-middle-con}
\end{align}
where (a) is due to the Markov chain $U-X_s-Y$ (cf.~\eqref{long-mk}).

The primary technical challenge in this proof is to find the optimal lower bound on the term $h(Y\vert{}U) - h(X_s\vert{}U)$ in the leakage rate. In what follows, we address this challenge by making use of a relation between the MMSE and  Fisher information.

For any pair of jointly distributed random variables $(A,V)$
such that $A$ has a differentiable conditional density given $V$,
the conditional Fisher information of $A$ is defined by
$
\mathbb{J}(A|V) = \mathbb{E}\left[\left(\frac{\partial\ln f_{A|V}(a|v)}{\partial a}\right)^2\right],
$
where the expectation is taken over $(A,V)$ \cite[Def. 1]{ekrem-2014}.

\begin{Lemma}[\hspace{-0.1mm}{\cite[Lemma 6]{hassan-2024}}] \label{hassan-lm6}
Let $A$ be a real-valued random variable with finite second moment,
$V$ an arbitrary random variable, and
$N' \sim \mathcal{N}(0,t)$ with $t>0$ independent of $(A,V)$.
Let $B=A+N'$. Then, it holds that
\begin{align*}
    \MMSE(A|V,B) = t - t^2\mathbb{J}(A + N'|V).
\end{align*}
\end{Lemma}

\begin{Lemma}[\hspace{-0.1mm}{\cite[Lemma 18]{Ekrem-2013}}] \label{key-lemma-1} Let $A, V, \tilde{N}, \bar{N}$ be random variables such that $(A,V)$ and $(\tilde{N}, \bar{N})$ are independent. Moreover, suppose $\tilde{N}  \sim  \mathcal{N}(0,\tilde{\sigma}^2)$ and $\bar{N} \sim \mathcal{N}(0,\bar{\sigma}^2)$ with  $0 < \tilde{\sigma}^2 \le \bar{\sigma}^2$. Then, it holds that
\begin{align*}
    \mathbb{J}(A + \bar{N}|{V})^{-1} - \bar{\sigma}^2 \ge \mathbb{J}(A + \tilde{N}|{V})^{-1} - \tilde{\sigma}^2.
\end{align*}
\end{Lemma}

\begin{Lemma}[\hspace{-0.1mm}{\cite[Lemma 16]{ekrem-2014}}] \label{lm-int}
Let $A$ be a real-valued random variable with finite second moment,
$V$ an arbitrary random variable, and
$N \sim \mathcal{N}(0,t)$ with $t>0$ independent of $(A,V)$.
It holds that
\begin{align*}
\frac{d}{dt} h(A + N|V) = \frac{1}{2\ln 2} \mathbb{J}(A + N|V).
\end{align*}
\end{Lemma}

Then, observe that in \eqref{xseq111}, it holds that
$0 < \MMSE(X|U,Y) \le \MMSE(X|Y) = (1-\rho^2_{XY})(\sigma^2_{X_s} + \sigma^2_{\Tilde{N}_X})$.
Therefore, there must exist an $\alpha \in (0,1]$ such that
\begin{align}
\MMSE(X|U,Y) = \frac{\alpha (1-\rho^2_{XY})(\sigma^2_{X_s} + \sigma^2_{\Tilde{N}_X})}{\alpha\rho^2_{XY} + 1-\rho^2_{XY}}. \label{mmse-setting}
\end{align}

For brevity, we define the following values
\begin{align}
t_1 = (1-\nu)\sigma^2_{X_s}/\nu^2~~{\rm and}~~t_2 = ((1-\nu)\sigma^2_{X_s}+ \sigma^2_{\Tilde{N}_Y})/\nu^2. \label{t1t2-def}
\end{align}
Note that $t_2 > \MMSE(X|U,Y)$ regardless of the value of $\alpha$.

\begin{Lemma} \label{lm-777} Let $N \sim \mathcal{N}(0,t)$ for $0 < t \le t_2$ be independent of $(X,U)$. Then, we have
\begin{align*}
    \mathbb{J}(X + N|U) \ge \frac{1}{t + \alpha(\sigma^2_{X_s} + \sigma^2_{\Tilde{N}_X})}.
\end{align*}
\end{Lemma}
\begin{proof}
Let $N_2 = 1/\nu N'_Y \sim \mathcal{N}(0,t_2)$. In Lemma \ref{hassan-lm6}, substituting $A \leftarrow X$, $N' \leftarrow N_2$, and $B \leftarrow (1/\nu) Y$ gives
\begin{align}
\mathbb{J}(X + N_2|U) = (t_2-\MMSE(X|U,Y))/t^2_2, \label{jxn2u}
\end{align}
where we use the fact that $\MMSE(X|U,(1/\nu)Y) = \MMSE(X|U,Y)$ as $1/\nu$ is constant. Moreover,  Lemma \ref{key-lemma-1} with $A \leftarrow X$, $\bar{N} \leftarrow N_2$, and $\tilde{N} \leftarrow N$ yields
\begin{align*}
\mathbb{J}(X + N_2|U)^{-1} - t_2 \ge \mathbb{J}(X + N|U)^{-1} - t,
\end{align*}
which implies that
\begin{align*}
    \mathbb{J}(X + N|U) &\ge \frac{1}{\mathbb{J}(X + N_2|U)^{-1} - t_2 + t} \nonumber \displaybreak[0] \\
    &\overset{\rm (a)}=\frac{1}{t + \frac{t_2\MMSE(X|U,Y)}{t_2 - \MMSE(X|U,Y)}} \nonumber \displaybreak[0] \\
    &\overset{\rm (b)}=\frac{1}{t + \alpha(\sigma^2_{X_s} + \sigma^2_{\Tilde{N}_X})},
\end{align*}
where (a) is due to \eqref{jxn2u} and (b) follows from \eqref{mmse-setting} and \eqref{t1t2-def}.
\end{proof}

Applying Lemma \ref{lm-int}, we have
\begin{align}
    h(Y|U) - h(X_s|U) &\ge \frac{1}{2\ln 2}\int_{t_1}^{t_2}\mathbb{J}(X + N|U)dt \nonumber \\
    &\overset{\rm (a)}\ge \frac{1}{2\ln 2}\int_{t_1}^{t_2}\frac{1}{t + \alpha(\sigma^2_{X_s} + \sigma^2_{\Tilde{N}_X})}dt \nonumber \\
    &= \frac{1}{2}\log\Big(\frac{t_2 + \alpha(\sigma^2_{X_s} + \sigma^2_{\Tilde{N}_X})}{t_1 + \alpha(\sigma^2_{X_s} + \sigma^2_{\Tilde{N}_X})}\Big) \nonumber \\
    &= \frac{1}{2} \log \Big( \frac{(\alpha\rho^2_{XY} + 1 - \rho^2_{XY})(\sigma_{X_s}^2 + \sigma_{\tilde{N}_Y}^2)}{(\alpha\rho^2_{XX_s} + 1 - \rho^2_{XX_s})\sigma_{X_s}^2} \Big),  \label{yxsu-eq}
\end{align}
where (a) is due to Lemma \ref{lm-777}.

Finally, the converse proof is completed by substituting \eqref{mmse-setting} and \eqref{yxsu-eq} into \eqref{d-middle-con}, \eqref{r-middle-con}, and \eqref{l-middle-con}.
\qed

\section{Proof of Theorem~\ref{inner-bound}} \label{app-b-a}
Fix the test channels $P_{U_i|X_{\Pi}}$ for $i \in \I$ such that $H(F|U_{\I},Y) = 0$. For small enough $\delta > 0$, we define
\begin{align}
R_i &= I(X_{\Pi};U_i|U_{[1:i-1]},Y) + {2\delta},~\forall i \in \I, \label{41ri-eqeq} \\
L &= I(X_{\S};U_{\I}|Y) + \delta. \label{41l-eqeq}
\end{align}

\subsection{Contra-Polymatroid Properties of the Achievable Rate Region} \label{appxc-a}
In what follows, we demonstrate that showing that the corner point with rates defined in \eqref{41ri-eqeq} and \eqref{41l-eqeq} is achievable suffices to establish the achievability of any tuple $((R_i)_{i \in \I},L)$ in \eqref{inner-ll}. To prove this, we state a series of definitions and lemmas.
\begin{Definition}[\hspace{-0.01mm}\cite{Edmonds2003}] \label{def-333}
Define the set  function $f:2^{\I} \rightarrow \mathbb{R}$. Then,
    \begin{itemize}
        \item[{\rm 1)}] $f$ is submodular if $f(\mathcal{A}\cup\mathcal{B}) + f(\mathcal{A}\cap\mathcal{B}) \le f(\mathcal{A}) + f(\mathcal{B})$ for $\mathcal{A}, \mathcal{B} \subseteq \I$.
        \item[{\rm 2)}] $f$ is supermodular if $-f$ is submodular.
    \end{itemize}
\end{Definition}
\begin{Definition}[\hspace{-0.01mm}\cite{Edmonds2003}] \label{def-444}
Define the set function $f:2^{\I} \rightarrow \mathbb{R}$,  the polyhedron $\mathcal{C}(f) \triangleq \{(R_i)_{i \in \I} \in \mathbb{R}^{\kappa}_+: R_{\A} \ge f(\mathcal{A}), \mathcal{A} \subseteq \I\}$ is a contra-polymatroid if
    \begin{itemize}
        \item[{\rm 1)}] $f$ is normalized, i.e., $f(\emptyset) = 0$,
        \item[{\rm 2)}] $f$ is non-decreasing, that is, $f(\mathcal{A}) \le f(\mathcal{B})$ for $\mathcal{A}, \mathcal{B} \subseteq \I$ with $\mathcal{A} \subseteq \mathcal{B}$,
        \item[{\rm 3)}] $f$ is supermodular.
    \end{itemize}
\end{Definition}

\begin{Lemma} \label{relabel-lemma}
For a given joint distribution $P_{U_{\I}X_{\P}Y} = P_{X_{\P}Y}\prod_{i=1}^{\kappa}P_{U_i|X_{\Pi}}$ and the set  function $f_{P_{U_{\I}X_{\P}Y}}:2^{\I} \rightarrow \mathbb{R}$, we have
\begin{itemize}
    \item[{\rm 1)}] For any $\mathcal{A} \subseteq \I$,
    \begin{align*}
    \mathcal{C}(f_{P_{U_{\I}X_{\P}Y}}) = \{&(R_i)_{i \in \I} \in \mathbb{R}^{\kappa}_+: \nonumber R_{\A} \ge f_{P_{U_{\I}X_{\P}Y}}(\mathcal{A}),~~{\A \subseteq \I}\},
    \end{align*}
    where $f_{P_{U_{\I}X_{\P}Y}}(\mathcal{A}) = I(X_{\P_{\A}};U_{\mathcal{A}}|U_{\mathcal{A}^c},Y)$, is a contra-polymatroid, with $\A^c\triangleq \I\setminus\A$.
    \item[{\rm 2)}] Let ${\rm Sym }(\I)$ represent the symmetric group on the set $\I$. Then, the dominant face of $\mathcal{C}(f_{P_{U_{\I}X_{\P}Y}})$, denoted by $\mathcal{D}(f_{P_{U_{\I}X_{\P}Y}})$, has the following property:
    \begin{align*}
        \mathcal{D}(f_{P_{U_{\I}X_{\P}Y}}) = {{\rm Conv}(\{(D^{\pi}_1,\dots,D^\pi_{\kappa}): \pi \in {\rm Sym }(\I)\})}, \label{26-26}
    \end{align*}
    where the vertices are given as
    \begin{align*}
        {D^{\pi}_{\pi(i)}} = I(X_{\P_{\pi(i)}};U_{\pi(i)}|U_{\pi(1:i-1)},Y)
    \end{align*}
    for $i \in \I$ and the notation $\pi(1:i-1) = \{\pi(j): j \in [1:i-1]\}$ with $U_{\pi(1:0)} = \emptyset$.
\end{itemize}
\end{Lemma}
\begin{proof}
First, we prove 1) by showing that the function $f_{P_{U_{\I}X_{\P}Y}}(\mathcal{A}) = I(X_{\P_{\A}};U_{\mathcal{A}}|U_{\mathcal{A}^c},Y)$, $\forall\A \subseteq \I$, satisfies conditions 1), 2), and 3) in Definition \ref{def-444}.

When $\mathcal{A} = \emptyset$, the function $f_{P_{U_{\I}X_{\P}Y}}(\mathcal{A}) = 0$, meaning it satisfies condition 1) in Definition \ref{def-444}. For any $\mathcal{A}, \mathcal{B} \subseteq \I$ with $\mathcal{A} \subseteq \mathcal{B}$, we have
\begin{align*}
    f_{P_{U_{\I}X_{\P}Y}}(\mathcal{B}) &= I(X_{\P_{\B}};U_{\mathcal{B}}|U_{\mathcal{B}^c},Y) \nonumber \displaybreak[0]\\
    &= I(X_{\P_{\B}};U_{\mathcal{B}\backslash \mathcal{A}}|U_{\mathcal{B}^c},Y) + I(X_{\P_{\B}};U_{\mathcal{A}}|U_{\mathcal{B}^c},U_{\mathcal{B}\backslash \mathcal{A}},Y) \nonumber \displaybreak[0]\\
    &\ge I(X_{\P_{\B}};U_{\mathcal{A}}|U_{\mathcal{B}^c},U_{\mathcal{B}\backslash \mathcal{A}},Y) \nonumber \displaybreak[0]\\
    &\overset{\rm (a)}= I(X_{\P_{\B}};U_{\mathcal{A}}|U_{\mathcal{A}^c},Y) \nonumber \displaybreak[0]\\
    &\ge I(X_{\P_{\A}};U_{\mathcal{A}}|U_{\mathcal{A}^c},Y) \nonumber \displaybreak[0]\\
    &= f_{P_{U_{\I}X_{\P}Y}}(\mathcal{A}),
\end{align*}
where (a) holds due to the relation $\mathcal{B}^c\cup (\mathcal{B}\backslash \mathcal{A}) = \mathcal{A}^c$. Hence, 2) in Definition \ref{def-444} is also satisfied. Finally, we show that the function $f_{P_{U_{\I}X_{\P}Y}}$ is supermodular. Note that for any set $\mathcal{A} \subseteq \I$, we have
\begin{align*}
I(X_{\P_{\A}};U_{\mathcal{A}}|U_{\mathcal{A}^c},Y) &\overset{\rm (a)}= I(X_{\P_{\I}};U_{\mathcal{A}}|U_{\mathcal{A}^c},Y) \nonumber \\
&\overset{\rm (b)}= I(X_{\P};U_{\I}|Y) - I(X_{\P};U_{\mathcal{A}^c}|Y),
\end{align*}
where (a) follows from the Markov chain $U_{\mathcal{A}}-(X_{\P_{\A}},U_{\mathcal{A}^c},Y)-X_{\P_{\A^c}}$, and (b) holds due to $\P = \P_{\I}$ and $\I = \A \cup \A^c$. Therefore, it suffices to show that the set function {$f':2^{\I} \rightarrow \mathbb{R}$ with $f'(\mathcal{B}) = I(X_{\P};U_{\mathcal{B}}|Y)$ for any $\mathcal{B} \subseteq \I$} is a submodular function by Definition \ref{def-333} since the term $I(X_{\P};U_{\I}|Y)$   does not depend on $\mathcal{A}$. For $\mathcal{A}, \mathcal{B} \subseteq \I$, we have
\[
\begin{aligned}
    f'(\mathcal{A}\cup\mathcal{B}) + f'(\mathcal{A}\cap\mathcal{B})
    &=I(X_{\P};U_{\mathcal{A}\cup\mathcal{B}}|Y) + I(X_{\P};U_{\mathcal{A}\cap\mathcal{B}}|Y) \\
    &= I(X_{\P};U_{\mathcal{A}}|Y) + I(X_{\P};U_{\mathcal{B}\backslash\mathcal{A}}|U_{\mathcal{A}},Y) + I(X_{\P};U_{\mathcal{B}}|Y) - I(X_{\P};U_{\mathcal{B}\backslash \mathcal{A}}|U_{\mathcal{A}\cap\mathcal{B}},Y) \\
    &= I(X_{\P};U_{\mathcal{A}}|Y) + I(X_{\P};U_{\mathcal{B}}|Y)  + H(U_{\mathcal{B}\backslash\mathcal{A}}|U_{\mathcal{A}},Y) - H(U_{\mathcal{B}\backslash \mathcal{A}}|U_{\mathcal{B}\cap\mathcal{A}},Y) \\
    &\le I(X_{\P};U_{\mathcal{A}}|Y) + I(X_{\P};U_{\mathcal{B}}|Y) \\
    &= f'(\mathcal{A}) + f'(\mathcal{B}).
\end{aligned}
\]

Next, we prove the statement 2). By 1) and \cite{Edmonds2003}, the vertices of the dominant faces are determined as
\begin{align*}
    {D^{\pi}_{\pi(i)}} &= f_{P_{U_{\I}X_{\P}Y}}(\pi(i:\kappa)) - f_{P_{U_{\I}X_{\P}Y}}(\pi(i+1:\kappa)), \quad \forall \pi \in {\rm Sym(\I)}, \forall i \in \I.
\end{align*}
Then, we have
\begin{align*}
    {D^{\pi}_{\pi(i)}} &= I(X_{\P_{\pi(i:\kappa)}};U_{\pi(i:\kappa)}|U_{\pi(i:\kappa)^c},Y) \nonumber - I(X_{\P_{\pi(i+1:\kappa)}};U_{\pi(i+1:\kappa)}|U_{\pi(i+1:\kappa)^c},Y) \nonumber \displaybreak[0] \\
    &= I(X_{\P_{\pi(i:\kappa)}};U_{\pi(i)}|U_{\pi(i:\kappa)^c},Y) \nonumber  + I(X_{\P_{\pi(i:\kappa)}};U_{\pi(i+1:\kappa)}|U_{\pi(i+1:\kappa)^c},Y) \nonumber  \\
    &~~~~ - I(X_{\P_{\pi(i+1:\kappa)}};U_{\pi(i+1:\kappa)}|U_{\pi(i+1:\kappa)^c},Y) \nonumber \displaybreak[0] \\
    &= I(X_{\P_{\pi(i:\kappa)}};U_{\pi(i)}|U_{\pi(i:\kappa)^c},Y) \nonumber + I(X_{\P_{\pi(i)}};U_{\pi(i+1:\kappa)}|X_{\P_{\pi(i+1:\kappa)}},U_{\pi(i+1:\kappa)^c},Y) \nonumber \displaybreak[0] \\
    &= I(X_{\P_{\pi(i:\kappa)}};U_{\pi(i)}|U_{\pi(1:i-1)},Y) \nonumber  + I(X_{\P_{\pi(i)}};U_{\pi(i+1:\kappa)}|X_{\P_{\pi(i+1:\kappa)}},U_{\pi(1:i)},Y) \nonumber \displaybreak[0] \\
    &\overset{\rm (a)}= I(X_{\P_{\pi(i)}};U_{\pi(i)}|U_{\pi(1:i-1)},Y) \nonumber \displaybreak[0] +I(X_{\P_{\pi(i+1:\kappa)}};U_{\pi(i)}|X_{\P_{\pi(i)}},U_{\pi(1:i-1)},Y) \nonumber \displaybreak[0] \\
    &\overset{\rm (b)}= I(X_{\P_{\pi(i)}};U_{\pi(i)}|U_{\pi(1:i-1)},Y),
\end{align*}
where (a) holds because $I(X_{\P_{\pi(i)}};U_{\pi(i+1:\kappa)}|X_{\P_{\pi(i+1:\kappa)}},U_{\pi(1:i)},Y) = 0$ because the mutual information is non-negative and we have $I(X_{\P_{\pi(i)}};U_{{\pi(i+1:\kappa)}}|X_{\P_{\pi(i+1:\kappa)}},U_{\pi(1:i)},Y) \le I(X_{\P_{\pi(i)}},U_{\pi(1:i)};U_{{\pi(i+1:\kappa)}}|X_{\P_{\pi(i+1:\kappa)}},Y) = 0$ by the Markov chain
$U_{{\pi(i+1:\kappa)}}-(X_{\P_{{\pi(i+1:\kappa)}}},Y)-(X_{\P_{\pi(i)}},U_{\pi(1:i)})$ and (b) is due to $I(X_{\P_{\pi(i+1:\kappa)}};U_{\pi(i)}|X_{\P_{\pi(i)}},U_{\pi(1:i-1)},Y) = 0$ as $I(X_{\P_{\pi(i+1:\kappa)}};U_{\pi(i)}|X_{\P_{\pi(i)}},U_{\pi(1:i-1)},Y) \le I(X_{\P_{\pi(i+1:\kappa)}},U_{\pi(1:i-1)};U_{\pi(i)}|X_{\P_{\pi(i)}},Y) = 0$ by the Markov chain
$U_{\pi(i)}-(X_{\P_{\pi(i)}},Y)-(X_{\P_{\pi(i+1:\kappa)}},U_{\pi(i+1:\kappa)})$.
\end{proof}

\begin{Lemma} \label{lemma-achiev} To prove the achievability of Theorem~\ref{inner-bound}, it is equivalent to show the achievability of the corner point with $(R_i)_{i \in \I}$ and $L$ defined in \eqref{41ri-eqeq} and \eqref{41l-eqeq}.
\end{Lemma}
\begin{proof}
Lemma~\ref{relabel-lemma} demonstrates that proving the achievability of $\mathcal{C}(f_{P_{U_{\I}X_{\P}Y}})$ is equivalent to showing the achievability of the tuple $(R_i)_{i \in \I}$ defined in \eqref{41ri-eqeq} as this tuple represents a corner point on the dominant face. This claim holds because other corner points on the dominant face can be achieved using the same proof with a relabeling of the DB indices, and any arbitrary point on the dominant face can be obtained through time-sharing among all the corner points. Furthermore, since the constraint \eqref{pri-def2} is preserved under time-sharing (remains unchanged for all corner points), the achievability of the tuple $((R_i)_{i \in \I},L)$ in \eqref{41ri-eqeq} and \eqref{41l-eqeq} implies the achievability of Theorem~\ref{inner-bound}.
\end{proof}

\subsection{Coding Scheme} \label{appx-c-b}

We now proceed to the proof of Theorem~\ref{inner-bound}, leveraging the OSRB technique \cite{Yassaee-tit2014}.

\subsubsection{Random Binning Scheme} Let $(U^n_{\I},X^n_{\P},X^n_{\S},Y^n)$ be i.i.d. sequences generated from the joint distribution $P_{X_{\P}X_{\S}Y}P_{U_{\I}|X_{\P}}$. {For $i \in \I$, define a random binning assignment $B_i: \mathcal{U}^n_i \to [1:2^{nR_i}]\times [1:2^{n\Tilde{R_i}}]$, where a sequence $U^n_i$ is uniformly and independently assigned to two bin indices $M_i$ and $C_i$ with rates $R_i$ and $\Tilde{R}_i$, respectively. The rate with $\Tilde{R_i}$ will be specified later}. The Slepian--Wolf (SW) decoder is used to estimate each $U^n_i$ and the decoding steps are as follows: Seeing $(M_1,C_1,Y^n)$, the decoder reconstructs $U^n_1$ as $\hat{U}^n_1$, and then it decodes $U^n_2$ as $\hat{U}^n_2$ from $(M_2,C_2,Y^n,\hat{U}^n_1)$. The procedure is successively repeated $\kappa$ times, where the sequence $U^n_i$ is reconstructed as $\hat{U}^n_i$ from $(M_i,C_i,Y^n,\hat{U}^n_{[1:i-1]})$ for $i \in \I$. Let $P(\cdot)$ denote the joint distribution induced by this random binning, given by
\begin{align*}
P_{M_{\I}C_{\I}U^n_{\I}X^n_{\P}X^n_{\S}Y^n\hat{U}^n_\I} &= P_{X^n_{\P}X^n_{\S}Y^n}\prod_{i=1}^{\kappa}P_{U^n_i|X^n_{\Pi}}P_{M_iC_i|U^n_i}P^{\rm SW}_{\hat{U}^n_i|M_iC_iY^n\hat{U}^n_{[1:i-1]}} \nonumber \\
&\overset{\rm (a)}= P_{X^n_{\P}X^n_{\S}Y^n}\prod_{i=1}^{\kappa}P_{U^n_i|X^n_{\Pi}}P_{M_i|U^n_i}P_{C_i|U^n_i}P^{\rm SW}_{\hat{U}^n_i|M_iC_iY^n\hat{U}^n_{[1:i-1]}} \nonumber \\
&\overset{\rm (b)}= P_{X^n_{\P}X^n_{\S}Y^n}\prod_{i=1}^{\kappa}P_{U^n_i|X^n_{\Pi}}P_{C_i|U^n_iX^n_{\Pi}}P_{M_i|U^n_i}P^{\rm SW}_{\hat{U}^n_i|M_iC_iY^n\hat{U}^n_{[1:i-1]}} \nonumber \\
&=P_{X^n_{\P}X^n_{\S}Y^n}\prod_{i=1}^{\kappa}P_{C_i|X^n_{\Pi}}P_{U^n_i|C_iX^n_{\Pi}}P_{M_i|U^n_i}P^{\rm SW}_{\hat{U}^n_i|M_iC_iY^n\hat{U}^n_{[1:i-1]}},
\end{align*}
where (a) holds because $M_i$ and $C_i$ are independent given $U^n_i$ and (b) holds by the Markov chain $C_i -U^n_i-X^n_{\Pi}$.

\subsubsection{Common-Randomness-Aided Coding Scheme} Suppose that the $i$-th encoder has access to the common randomness $C_i, i \in \I,$ uniformly distributed over $[1:2^{n\Tilde{R}_i}]$. The decoder also has knowledge of all $C_i, i \in \I$. The encoder generates $U^n_i$ according to $P_{U^n_i|C_iX^n_{\Pi}}$, then transmits message $M_i$ generated according to $P_{M_i|U^n_i}$. The SW decoder is used to reconstruct an estimate of $U^n_i$ as $\hat{U}^n_i$, and the decoding procedure is the same as explained in the random binning scheme. The joint distribution induced by this coding scheme, denoted by $\Tilde{P}(\cdot)$, is
\begin{align*}
\Tilde{P}_{M_{\I}C_{\I}U^n_{\I}X^n_{\P}X^n_{\S}Y^n\hat{U}^n_\I}
= P_{X^n_{\P}X^n_{\S}Y^n} \prod_{i=1}^{\kappa}P^{\rm U}_{C_i} P_{U^n_i|C_iX^n_{\P_i}}P_{M_i|U^n_i}P^{\rm SW}_{\hat{U}^n_i|M_iC_iY^n\hat{U}^n_{[1:i-1]}},
\end{align*}
where $P^{\rm U}_{C_i}$ is the uniform distribution over the set $[1:2^{n\Tilde{R}_i}]$.

{In the following, we write $P^{b_{\I}}$ and $\tilde{P}^{b_{\I}}$ for the laws induced by the binning scheme and the common-randomness-aided coding scheme with respect to a fixed binning assignment $B_{\I} = b_{\I}$}.

\begin{Lemma}[\hspace{-0.01mm}{\cite[Th.~1]{Yassaee-tit2014}}] \label{lm-6} If {$\Tilde{R}_i < H(U_i|X_{\Pi})$}, the index $C_i$ is almost independent of $X^n_{\Pi}$ for $i \in \I$, i.e.,
\begin{align*}
    \lim_{n \to \infty} {\E_{B_\I}} \left[\V(P^{B_\I}_{C_iX^n_{\Pi}},P^{\rm U}_{C_i}P_{X^n_{\Pi}})\right] = 0,
\end{align*}
where {$\E_{B_\I}[\cdot]$} is  expectation over the random binning assignments, $\V(P_X,Q_X)\!=\! \frac{1}{2}\sum_{x \in \mathcal{X}}|P_X(x)-Q_X(x)|$ is the total variation distance between two  probability distributions $P_X$ and $Q_X$ defined on  $\mathcal{X}$.
\end{Lemma}

\begin{Lemma}[\hspace{-0.01mm}{\cite[Lemma~1]{Yassaee-tit2014}}] \label{lm-7} When  {$R_i + \tilde{R}_i \ge H(U_i|U_{[1:i-1]},Y) + \delta$} for $i \in \I$, the SW decoder can reliably estimate $U^n_i$ from {$(M_i,C_i,Y^n,U^n_{[1:i-1]})$}, i.e.,
\begin{align}
    \lim_{n \to \infty} {\E_{B_\I}}\left[\V(P^{B_\I}_{M_\I C_\I U^n_\I X^n_\P {X^n_\S}Y^n\hat{U}^n_\I},P^{B_\I}_{M_\I C_\I U^n_\I X^n_\P X^n_\S Y^n}\mathds{1}\{\hat{U}^n_\I = U^n_\I\})\right] = 0. \label{eqeq-70}
\end{align}
If $H(U_i|X_{\mathcal{P}_i}) = 0$, we set $\Tilde{R}_i = 0$, in which case the sum-rate condition reduces to {$R_i \ge H(U_i|U_{[1:i-1]}, Y) + \delta$}.
\end{Lemma}

By Lemma \ref{lm-6}, if we select {$\Tilde{R}_i = \max\{H(U_i|X_{\Pi}) - \delta,0\}$}, the index $C_i$ is almost independent of $X^n_{\P_i}$ (and also $X^n_{\S_i}$ and $Y^n$). Hence, we have
\begin{align}
\lim_{n \rightarrow \infty}&{\E_{B_\I}} \Big[\V(\Tilde{P}^{B_\I}_{M_{\I}C_{\I}U^n_{\I}X^n_{\P}X^n_{\S}Y^n\hat{U}^n_{\I}},P^{B_\I}_{M_{\I}C_{\I}U^n_{\I}X^n_{\P}X^n_{\S}Y^n\hat{U}^n_{\I}})\Big] \nonumber \displaybreak[0] \nonumber \displaybreak[0]  \\
&\overset{\rm (a)}= \lim_{n \rightarrow \infty}{\E_{B_\I}} \Big[\V(\Tilde{P}^{B_\I}_{M_{\I}C_{\I}U^n_{\I}X^n_{\P}X^n_{\S}Y^n},P^{B_\I}_{M_{\I}C_{\I}U^n_{\I}X^n_{\P}X^n_{\S}Y^n})\Big]  \nonumber \displaybreak[0]  \\
&= 0, \label{rbcs-joint}
\end{align}
where (a) follows because the same SW decoder is used in both $P^{b_\I}(\cdot)$ and $\Tilde{P}^{b_\I}(\cdot)$. By Lemma \ref{lm-7}, the SW decoder can reliably estimate $U^n_i$ from $(M_i,C_i,Y^n,\hat{U}^n_{[1:i-1]})$ by choosing $R_i + \Tilde{R}_i = H(U_i|U_{[1:i-1]},Y) + \delta$. By \eqref{eqeq-70}, \eqref{rbcs-joint}, and the triangle inequality, we have
\begin{align}
    \lim_{n \rightarrow \infty}{\E_{B_\I}} \Big[\V(\Tilde{P}^{B_\I}_{M_{\I}C_{\I}U^n_{\I}X^n_{\P}X^n_{\S}Y^n\hat{U}^n_{\I}},P^{B_\I}_{M_{\I}C_{\I}U^n_{\I}X^n_{\P}X^n_{\S}Y^n}\mathds{1}\{\hat{U}^n_\I = U^n_\I\})\Big] = 0. \label{p-bar-1}
\end{align}

\subsubsection{Coding Scheme Analysis}\mbox{}

\emph{Error Probability}: As we choose the test channel $P_{U_\I|X_\P}$ such that $H(F|U_{\I},Y) = 0$, there exists a deterministic function $g$ such that $g(u_{\I},y) = f(x_{\P},y)$ for all tuples $(u_{\I},x_{\P},y)$ with $P_{U_{\I}X_{\P}Y}(u_{\I},x_{\P},y) > 0$. For i.i.d. samples, $g_n$ recovers $F^n$ exactly from $(U_\I^n,Y^n)$ with probability one. Finally, the decoder computes $\hat{F}^n=g^n(\hat{U}^n_{\I},Y^n)$. As shown in \eqref{p-bar-1}, the probability in which $\hat{U}^n_{\I} = U^n_{\I}$ approaches one, and therefore we have
$
\lim_{n \rightarrow \infty}\mathbb{P}[\hat{F}^n = F^n] =1, 
$
which in turn ensures that
\begin{align}
    \lim_{n \rightarrow \infty}\mathbb{P}[\hat{F}^n \neq F^n] = 0. \label{error-pro}
\end{align}

\emph{Communication Rate}: For $i \in \I$, we have
\begin{align}
    \frac{1}{n}\log|\M_i| &= R_i \nonumber \\ 
    &= H(U_i|U_{[1:i-1]},Y) + \delta - \Tilde{R}_i \nonumber \\
    &= H(U_i|U_{[1:i-1]},Y) - H(U_i|X_{\Pi}) + 2\delta \nonumber \\
    &\overset{\rm (a)}= H(U_i|U_{[1:i-1]},Y) - H(U_i|X_{\Pi},U_{[1:i-1]},Y) + 2\delta \nonumber \\
    &= I(X_{\Pi};U_i|U_{[1:i-1]},Y) + 2\delta, \label{r-111}
\end{align}
where (a) follows from the Markov chain $U_i-X_{\Pi}-(U_{[1:i-1]},Y)$.

\emph{Leakage Rate}:~~~Let $I(\cdot;\cdot)$ and $I_{\Tilde{P}^{b_\I}}(\cdot;\cdot)$ denote the mutual information calculated with respect to the laws $P^{b_\I}(\cdot)$ and $\Tilde{P}^{b_\I}(\cdot)$, respectively. Since the decoder also observes the common randomness, the {left-hand side} of \eqref{pri-def2} can be evaluated as
\begin{align}
{\E_{B_\I}}\left[I_{\Tilde{P}^{B_\I}}(X^n_{\S};M_{\I}|Y^n,C_\I)\right]
&\le \E_{B_\I}\left[I_{\Tilde{P}^{B_\I}}(X^n_{\S};U^n_\I,M_\I|Y^n,C_\I)\right]\nonumber \displaybreak[0] \\
&\overset{\rm (a)}= \E_{B_\I}\left[I_{\Tilde{P}^{B_\I}}(X^n_{\S};U^n_\I|Y^n,C_\I)\right] \nonumber \displaybreak[0] \\
&\overset{\rm (b)}= \E_{B_\I}\left[I(X^n_{\S};U^n_\I|Y^n,C_\I)\right] + o(n) \nonumber \displaybreak[0] \\
&=  \E_{B_\I}\left[I(X^n_{\S};U^n_\I,C_\I|Y^n) - I(X^n_{\S};C_\I|Y^n)\right]  + o(n) \nonumber \displaybreak[0] \\
&\le  \E_{B_\I}\left[I(X^n_{\S};U^n_\I,C_\I|Y^n)\right]  + o(n) \nonumber \displaybreak[0] \\
&\overset{\rm (c)}=  \E_{B_\I}\left[I(X^n_{\S};U^n_\I|Y^n)\right]  + o(n) \nonumber \displaybreak[0] \\
&\overset{\rm (d)}= nI(X_{\S};U_{\I}|Y) + o(n) \nonumber \displaybreak[0] \\
&\overset{\rm (e)}\le nL \label{l-111}
\end{align}
for large enough $n$, where (a) holds because $M_\I$ is generated from $U^n_\I$ in the coding scheme, (b) follows by applying \cite[Lemma 1]{csiszar1996} due to \eqref{rbcs-joint}, (c) holds as $C_\I$ is a function of $U^n_\I$ in the random binning scheme, (d) holds since $(X^n_{\S},U^n_{\I},Y^n)$ is i.i.d.~due to  the i.i.d. test-channel distribution in the binning scheme, and (e) follows from $L$ defined in \eqref{41l-eqeq}.

\subsubsection{Derandomization}\mbox{}
We now perform a simultaneous selection of the realizations for the random binning scheme and the common randomness.

For a realization pair $(b_\I, c_\I)$ of the binning assignment and common randomness, we define 
\begin{align}
\delta_n(b_\I,c_\I) &\triangleq \mathbb{P}\left[F^n \neq  \hat{F}^n|(B_\I,C_\I) = (b_\I,c_\I) \right], \label{delta-n-f}\\
\ell_n(b_\I,c_\I) &\triangleq \frac{1}{n} {I_{\tilde{P}^{b_\I}}} (X_{\S}^n; M_\I|Y^n,C_\I=c_\I), \label{ell-n-f}
\end{align}
where \eqref{delta-n-f} is the error probability, where the target function is not identical to its reconstruction $\hat{F}^n = g^n(\hat{U}^n_\I,Y^n)$, while \eqref{ell-n-f} quantifies the leakage rate. Taking expectations over $B_\I$ and $\tilde{P}^{b_\I}$, we have  $\epsilon_n \triangleq \mathbb{E}_{B_\I}\left[\mathbb{E}_{\tilde{P}^{B_\I}}[\delta_n(B_\I, C_\I)]\right] \longrightarrow 0 \quad \text{as } n \to \infty$ by  \eqref{error-pro} and $\mathbb{E}_{B_\I}\left[\mathbb{E}_{\tilde{P}^{B_\I}}[\ell_n(B_\I, C_\I)]\right] \le L $ by \eqref{l-111}. Now define a sequence $\mu_n \triangleq \sqrt{\epsilon_n} + n^{-1}$, which satisfies $\mu_n \to 0$ and $\frac{\epsilon_n}{\mu_n} \to 0$ as $n \to \infty$. By Markov's inequality, it holds that
\begin{align*}
\mathbb{P}\{ \delta_n(B_\I, C_\I) > \mu_n \} &\le \frac{\mathbb{E}_{B_\I}\left[\mathbb{E}_{\tilde{P}^{B_\I}}[\delta_n(B_\I, C_\I)]\right]}{\mu_n} = \frac{\epsilon_n}{\mu_n}, \\
\mathbb{P}\{ \ell_n(B_\I, C_\I) > L + \delta \} &\le \frac{\mathbb{E}_{B_\I}\left[\mathbb{E}_{\tilde{P}^{B_\I}}[\ell_n(B_\I, C_\I)]\right]}{L + \delta} \le \frac{L}{L + \delta}.
\end{align*}
Applying the union bound to the complement event yields
\begin{align*}
\mathbb{P}\left\{ \delta_n(B_\I, C_\I) \le \mu_n, \ell_n(B_\I, C_\I) \le L + \delta \right\} \ge 1 - \frac{\epsilon_n}{\mu_n} - \frac{L}{L + \delta} = \frac{\delta}{L + \delta} - \frac{\epsilon_n}{\mu_n} \overset{n \to \infty}\longrightarrow \frac{\delta}{L + \delta} > 0.
\end{align*}
Since this probability is strictly positive for sufficiently large $n$, there exists at least one realization pair $(b_{\mathcal{I}}, c_{\mathcal{I}})$ such  that
\begin{align*}
\mathbb{P}\left[\hat{F}^n \neq  F^n|(B_\I,C_\I) = (b_\I,c_\I)\right] &\le \mu_n, \\
\frac{1}{n}I_{\Tilde{P}^{b_\I}}(X^n_\S;M_\I|Y^n,C_\I=c_\I) &\le L + \delta.
\end{align*}
Note that the communication rates $R_i = \frac{1}{n}\log |\mathcal{M}_i|$ for $i \in \I$ depend on the pre-fixed bin cardinalities, ensuring that the selected realization pair $(b_\I, c_\I)$ does not affect the achievable rates. Finally, by taking $n \to \infty$ and $\delta \downarrow 0$, the corner point $((R_i)_{i\in\mathcal{I}},L)$ defined in \eqref{41ri-eqeq} and \eqref{41l-eqeq} is achieved.
\qed

\section{Proof of Theorem~\ref{outer-bound}} \label{app-b-b}

Suppose that the tuple $((R_i)_{i \in \I},L)$ is achievable under the constraints \eqref{rate-def2}, \eqref{pri-def2}, and \eqref{error-def1}. For $t \in [1:n]$ and $i \in \I$, we define random variables
$
U_{i,t} \triangleq (M_i,X^{t-1}_{\Si})
$
and
$Q_t \triangleq Y^{n \backslash t}$. This setting satisfies the Markov chain $(Q_t,U_{i,t}) - X_{\Pi,t}-(X_{\P\setminus \Pi,t},X_{\S,t},Y_t)$ for any $i \in \I$ since
\begin{align}
    P_{Q_tU_{i,t}X_{\P,t}X_{\S,t}Y_t} &= P_{Y^{n \setminus t}M_iX^{t-1}_{\Si}X_{\P,t}X_{\S,t}Y_t} \nonumber \displaybreak[0]\\
    &= \sum_{x^{n\setminus t}_{\Pi} \in \mathcal{X}^{n\setminus t}_{\Pi}}P_{Y^{n \setminus t}M_iX^{t-1}_{\Si}X^n_{\Pi,t}X_{\P\setminus \Pi,t}X_{\S,t}Y_t} \nonumber \displaybreak[0]\\
    &\overset{\rm (a)}= \sum_{x^{n\setminus t}_{\Pi} \in \mathcal{X}^{n\setminus t}_{\Pi}}\Big\{P_{Y^{n \setminus t}M_iX^{t-1}_{\Si}X^n_{\Pi}}\Big\}\cdot P_{X_{\P\setminus \Pi,t}X_{\S,t}Y_t|X_{\Pi,t}} \nonumber \displaybreak[0]\\
    &= P_{Y^{n \setminus t}M_iX^{t-1}_{\Si}X_{\Pi,t}}\cdot P_{X_{\P\setminus \Pi,t}X_{\S,t}Y_t|X_{\Pi,t}} \nonumber \displaybreak[0]\\
    &= P_{Q_tU_{i,t}X_{\Pi,t}}\cdot P_{X_{\P\setminus \Pi,t}X_{\S,t}Y_t|X_{\Pi,t}}, \label{mk-converse}
\end{align}
where (a) holds because $M_i$ is a function of $X^n_{\Pi}$ and $(X^n_{\P},X^n_{\S},Y^n)$ is an i.i.d.~sequence.

For any $t \in [1:n]$,  we have
\begin{align}
H(F_t|U_{\I,t},Y_t,Q_t)
&=
H(F_t|M_\I,X_\S^{t-1},Y^n) \nonumber \displaybreak[0]\\
&\overset{\rm (a)}{\le}
H(F_t|M_\I,Y^n) \nonumber \displaybreak[0]\\
&\overset{\rm (b)}{=}
H(F_t|M_\I,Y^n,\hat{F}^n) \nonumber \displaybreak[0]\\
&\overset{\rm (c)}{\le}
H(F_t|\hat{F}_t) \nonumber \displaybreak[0]\\
&\overset{\rm (d)}{\le}
H_b(\delta)+\delta\log(|\mathcal{F}|-1), \label{hfuy-ad}
\end{align}
where (a) and (c) hold because conditioning does not increase entropy, (b) holds by the Markov chain $F_t-(M_\I,Y^n)-\hat{F}^n$, and (d) holds by Fano's inequality  because
$\mathbb{P}[\hat{F}_t\neq F_t] \le \mathbb{P}[\hat{F}^n\neq F^n] \le \delta$. 

{\em Analysis of Communication Rate}:~~~From \eqref{rate-def2}, it follows that for any $\A \subseteq \I$,
\begin{align*}
    n(R_{\A} + |\A|\delta) &\ge \sum_{i \in \A}\log |\M_i| \nonumber \\
    &\ge H(M_{\A}) \nonumber \\
    &\ge I(X^n_{\P_{\A}};M_{\A}) \nonumber \displaybreak[0] \\
    & \ge I(X^n_{\P_{\A}};M_{\A}) - I(Y^n;M_{\A}) \nonumber \displaybreak[0]\\
    &\overset{\rm (a)}=I(X^n_{\P_{\A}};M_{\A}|Y^n) \nonumber \displaybreak[0] \\
    &=\sum_{t=1}^nI(X_{\P_{\A},t};M_{\A}|X^{t-1}_{\P_{\A}},Y^n) \nonumber \displaybreak[0] \\
    &\overset{\rm (b)}= \sum_{t=1}^nI(X_{\P_{\A},t};M_{\A}|X^{t-1}_{\P_{\A}},X^{t-1}_{\S_{\A}},Y^n) \nonumber \displaybreak[0] \\
    &\overset{\rm (c)}= \sum_{t=1}^nI(X_{\P_{\A},t};M_{\A},X^{t-1}_{\P_{\A}},X^{t-1}_{\S_{\A}}|Y_t,Q_t) \nonumber \\
    &\ge \sum_{t=1}^nI(X_{\P_{\A},t};M_{\A},X^{t-1}_{\S_{\A}}|Y_t,Q_t) \nonumber \displaybreak[0] \\
    &= \sum_{t=1}^nI(X_{\P_{\A},t};U_{\A,t}|Y_t,Q_t) \\
    &= \sum_{t=1}^n\{I(X_{\P_{\A},t},U_{\A^c,t};U_{\A,t}|Y_t,Q_t) - I(U_{\A^c,t};U_{\A,t}|X_{\P_{\A},t},Y_t,Q_t)\} \nonumber \displaybreak[0] \\
    &= \sum_{t=1}^n\Big\{I(X_{\P_{\A},t};U_{\A,t}|U_{\A^c,t},Y_t,Q_t) \nonumber \displaybreak[0] \\
    &~~~~+ I(U_{\A,t};U_{\A^c,t}|Y_t,Q_t) - I(U_{\A,t};U_{\A^c,t}|X_{\P_{\A},t},Y_t,Q_t)\Big\} \nonumber \displaybreak[0] \\
    &\ge \sum_{t=1}^n\Big\{I(X_{\P_{\A},t};U_{\A,t}|U_{\A^c,t},Y_t,Q_t) \nonumber \displaybreak[0] \\
    &~~~~+ [I(U_{\A,t};U_{\A^c,t}|Y_t,Q_t) - I(U_{\A,t};U_{\A^c,t}|X_{\P_{\A},t},Y_t,Q_t)]^-\Big\},
\end{align*}
where (a) is due to the Markov chain $M_{\A}-X^n_{\P_{\A}}-Y^n$, (b) follows from the Markov chain $X_{\P_{\A},t}-(M_{\A},X^{t-1}_{\P_{\A}},Y^n)-X^{t-1}_{\S_{\A}}$, and (c) is due to the i.i.d. property of $(X^n_{\P_{\A}},X^n_{\S_{\A}},Y^n)$.

{\em Analysis of Leakage Rate}:~~~From \eqref{pri-def2}, we have
\begin{align*}
    n(L + \delta) &\ge I(X^n_{\S};M_{\I}|Y^n) \nonumber \\
    &= \sum_{t=1}^nI(X_{\S,t};M_{\I}|X^{t-1}_{\S},Y^n) \nonumber \displaybreak[0] \\
    &\overset{\rm (a)}= \sum_{t=1}^nI(X_{\S,t};M_{\I},X^{t-1}_{\S}|Y^n) \nonumber \displaybreak[0] \\
    &= \sum_{t=1}^nI(X_{\S,t};M_{\I},X^{t-1}_{\S}|Y_t,Y^{n \setminus t}) \nonumber \displaybreak[0] \\
    &= \sum_{t=1}^nI(X_{\S,t};U_{\I,t}|Y_t,Q_t),
\end{align*}
where (a) follows because $X^n_{\S}$ and $Y^n$ are i.i.d.~sequences.

Introduce a time-sharing random variable $T$ uniformly distributed over $[1:n]$ and independent of other random variables, and define $Q \triangleq (Q_T,T)$, $(U_i \triangleq U_{i,T}, X_{\Pi} \triangleq X_{\Pi,T}, X_{\Si} \triangleq X_{\Si,T})$ for $i \in \I$, and $Y \triangleq Y_T$, so that the Markov chain $(Q,U_i)-X_{\P_i}-(X_{\P \backslash \P_i},X_{\S},Y)$ holds for $i \in \I$. Also, we define the function $F \triangleq F_T = f_T(X_\P,Y)$. The cardinality bounds of $U_i$ ($i \in \I$) and $Q$ are shown in Lemma \ref{CB-ll}. Finally, by letting $n \rightarrow \infty$ and $\delta \downarrow 0$,  any achievable rate tuple is included in~\eqref{outer-ll}.
\qed

\section{Proof of Theorem~\ref{th3-3-ll}} \label{app-c}

The achievability proof follows similarly to that of Theorem \ref{inner-bound} in Appendix~\ref{app-b-a}, and thus it is omitted. In this appendix, we only provide the converse proof.

Assume that the tuple $((R_i)_{i \in \I},L)$ is achievable with respect to \eqref{rate-def2}, \eqref{pri-def2}, and \eqref{error-def1}. Define $U_{i,t} \triangleq (M_i,X^{t-1}_{\S_i})$ and $Q_t \triangleq Y^{n \setminus t}$. Due to the independence between $(X_{\Si},X_{\Pi})$ and $(X_{\S_j},X_{\P_j})$ for $i,j \in \I$ with $i \neq j$ as well as $Y$, this choice of $U_{i,t}$ satisfies   \eqref{th6-mk}.

The condition $H(F_t|U_{\I,t},Y_t,Q_t) \le H_b(\delta)+\delta\log(|\mathcal{F}|-1)$ follows from the same argument as in \eqref{hfuy-ad}. The analyses of the communication and leakage rates are as follows.

\smallskip
{\em Analysis of Communication Rate}:~~~From \eqref{rate-def2}, we have
\begin{align*}
    n(R_i + \delta ) &\ge \log|\M_i| \nonumber \\
    &\ge H(M_i) \nonumber \displaybreak[0] \\
    &\ge I(X^n_{\Pi};M_i) \nonumber \displaybreak[0] \\
    &=\sum_{t=1}^nI(X_{\Pi,t};M_i|X^{t-1}_{\Pi}) \nonumber \displaybreak[0] \\
    &\overset{\rm (a)}=\sum_{t=1}^nI(X_{\Pi,t};M_i,X^{t-1}_{\Pi}) \nonumber \displaybreak[0] \\
    &\overset{\rm (b)}=\sum_{t=1}^nI(X_{\Pi,t};M_i,X^{t-1}_{\Pi},X^{t-1}_{\S_i},Y^{n \backslash t}) \nonumber \displaybreak[0] \\
    &\ge \sum_{t=1}^nI(X_{\Pi,t};M_i,X^{t-1}_{\S_i},Y^{n \backslash t}) \nonumber \displaybreak[0] \\
    &= \sum_{t=1}^nI(X_{\Pi,t};M_i,X^{t-1}_{\S_i}|Y^{n \backslash t}) \nonumber \displaybreak[0] \\
    &= \sum_{t=1}^nI(X_{\Pi,t};U_{i,t}|Q_t),
\end{align*}
where (a) holds because $X^n_{\P}$ is i.i.d. and (b) is due to the Markov chain $X_{\Pi,t}-(M_i,X^{t-1}_{\Pi})-(X^{t-1}_{\S_i},Y^{n \backslash t})$ as $Y^{n \backslash t}$ is independent of $(X^t_{\Pi},M_i,X^{t-1}_{\S_i})$.

\smallskip
{\em Analysis of Leakage Rate}:~~~From \eqref{pri-def2}, it follows that
\begin{align*}
    n(L + \delta) &\ge I(X^n_{\S};M_{\I}|Y^n) \nonumber \displaybreak[0]\\
    &= \sum_{t=1}^nI(X_{\S,t};M_{\I}|X^{t-1}_{\S},Y^n) \nonumber \displaybreak[0]\\
    &\overset{\rm (a)}= \sum_{t=1}^nI(X_{\S,t};M_{\I},X^{t-1}_{\S}|Y^n) \nonumber \displaybreak[0]\\
    &= \sum_{t=1}^nI(X_{\S,t};U_{\I,t}|Y_t,Q_t) \nonumber \displaybreak[0]\\
    &\overset{\rm (b)}= \sum_{t=1}^nI(X_{\S,t};U_{\I,t}|Q_t),
\end{align*}
where (a) is due to the i.i.d. property among symbols in $(X^n_{\S},Y^n)$ and (b) is due to the independence of $(X_{\S,t},X_{\P,t})$ and $Y_t$, and thus $(U_{\I,t},X_{\S,t})$ and $Y_t$ are also independent.

Let $T$ be a random variable uniformly distributed over $[1:n]$ and independent of other random variables. We define $Q \triangleq (Q_T,T)$, $U_i \triangleq U_{i,T},X_{\mathcal{P}_i} \triangleq X_{\mathcal{P}_i,T},X_{\mathcal{S}_i} \triangleq X_{\mathcal{S}_i,T}$, and $Y \triangleq Y_T$, which satisfies the Markov chain $(Q,U_i)-X_{\Pi}-(X_{\P\setminus \Pi},X_\S,Y)$ for $i \in \I$, and $F \triangleq F_T = f_T(X_{\P},Y)$. The cardinality bounds of $U_i$ for $i \in \I$ and $Q$ are shown in Lemma~\ref{CB-ll}. By letting $n \rightarrow \infty$ and $\delta \downarrow 0$, any achievable rate tuple is included in~\eqref{indep-eq-ll}.
\qed

\section{Proof of Lemma \ref{CB-ll}} \label{Appx-CB-ll}
\begin{enumerate}
\item[(a)] Fix $Q = q$, where $q \in \mathcal{Q}$, and the conditional distributions $P_{U_{i'}|X_{\P_{i'}}Q=q}$ for $i' \in \I\setminus\{i\}$. The right-hand side of the communication rates can be expanded as $I(X_{\mathcal{P}_{\mathcal{A}}};U_{\mathcal{A}}|U_{\mathcal{A}^c},Y,Q=q) = H(X_{\mathcal{P}_{\mathcal{A}}}|U_{\mathcal{A}^c},Y,Q=q)-H(X_{\mathcal{P}_{\mathcal{A}}}|U_{\mathcal{A}},U_{\mathcal{A}^c},Y,Q=q)$. Applying the support lemma \cite[Appx.~C]{GK} to the distribution $P_{X_{\P_i}|U_iQ=q}$, preserving the distribution $P_{X_{\P_i}}$ requires $|\mathcal{X}_{\mathcal{P}_i}|-1$ elements. In addition, preserving $H(X_{\mathcal{P}_{\mathcal{A}}}|U_{\mathcal{A}^c},Y,Q=q)$ and $H(X_{\mathcal{P}_{\mathcal{A}}}|U_{\mathcal{A}},U_{\mathcal{A}^c},Y,Q=q)$ across all non-empty subsets $\mathcal{A},\mathcal{A}^c \subseteq \I = [1:\kappa]$ needs $2^{\kappa}-1$ elements each with two additional elements to preserve $H(F|U_{\mathcal{I}},Y,Q=q)$ and $I(X_{\mathcal{S}};U_{\mathcal{I}}|Y,Q=q)$.
As a result, it holds that $|\mathcal{U}_i| \le |\X_{\P_i}| + 2^{\kappa+1} - 1$. Repeating this process for each encoder and other $q$'s yields $|\mathcal{U}_i| \le |\X_{\P_i}| + 2^{\kappa+1} - 1$ for $i \in \I$. 
\item[(b)] Unlike the previous proof of part (a) for the inner bound, it seems difficult to apply the support lemma to bound the cardinalities of the auxiliary random variables because we must account for the correlations among the auxiliary random variables, which prevents applying the support lemma to each individual test channel $P_{U_i|X_iQ}$ for $i \in \I$. To address this, we present an alternative proof that constructs replacement auxiliary variables $(\tilde{U}_\I,\tilde{Q})$ for the original $(U_\I,Q$) while preserving the Markov conditions, the reconstruction and rate constraints. The proof consists of the following steps.
\begin{itemize}
\item \emph{Step 1: Construction of the replacement auxiliary variables}. Let $\mathcal{F} \triangleq [0:m-1]$. Choose mutually independent random variables $K_1,\dots,K_{\kappa-1}$ uniform on $[0:m-1]$ and independent of all other random variables. Take $\tilde{Q}$ constant and define  $\tilde{U}_i \triangleq K_i$ for $i \in [1:\kappa-1]$ and $\tilde{U}_{\kappa} \triangleq \left(F - \sum_{i=1}^{\kappa - 1}K_i\right) \mod m$, which ensures that $F = \sum_{i=1}^{\kappa}\tilde{U}_i \mod m$. Therefore,  $H(F|\tilde{U}_\I,Y) = 0$.
\item \emph{Step 2: Verification of the Markov conditions in \eqref{outer-mk}}. For $i \in [1:\kappa-1]$, $\tilde{U}_i$ is uniform and independent of $(X_\P,X_\S,Y)$, hence, $P_{\tilde{U}_i|X_\P X_\S Y}(u_i|x_\P,x_\S,y)=\frac{1}{m}$. For the last auxiliary random variable, given a tuple $(x_\P,x_\S,y)$, the function $F=f(x_\P,y)$ is fixed, so we have
    $$P_{\tilde{U}_\kappa|X_\P,X_\S,Y}(u_\kappa|x_\P,x_\S,y) = \mathbb{P}\left[\sum_{i = 1}^{\kappa-1}K_i = f(x_\P,y)-u_\kappa \pmod m\right] = \frac{1}{m}.$$
    Consequently, for every $i \in \I$, 
    $
    P_{\tilde{U}_i|X_\P,X_\S,Y}(u_i|x_\P,x_\S,y) = \frac{1}{m} = P_{\tilde{U}_i|X_{\Pi}}(u_i|x_{\Pi}),
    $
    which proves that $\tilde{U}_i-X_{\Pi}-(X_{\P\setminus\Pi},X_\S,Y)$. Since $\tilde{Q}$ is constant, the Markov conditions in \eqref{outer-mk} hold.
\item \emph{Step 3: Verification of the communication-rate constraints for non-empty subsets $\A \subset \I$}. We first show that $\tilde{U}_{\mathcal{A}}$ is independent of the sources. For a fixed source realization $(x_\P,x_\S,y)$, each tuple satisfying $f(x_{\mathcal{P}}, y) = \sum_{i=1}^{\kappa} u_i \pmod m$ has a probability of $m^{-(\kappa - 1)}$. After fixing $u_{\mathcal{A}}$, there are $\kappa - |\mathcal{A}|$ coordinates left, in which all but one can be chosen freely. The last coordinate is determined by the required sum. Thus, there are $m^{\kappa - |\mathcal{A}| - 1}$ compatible completions, and $P_{\widetilde U_{\mathcal A}\mid X_{\mathcal P},X_{\mathcal S},Y}(u_{\mathcal{A}}|x_\P,x_\S,y) = \frac{m^{\kappa - |\mathcal{A}| - 1}}{m^{\kappa - 1}} = m^{-|\mathcal{A}|}$. This probability does not depend on $(x_\P,x_\S,y)$, therefore, $I(X_{\mathcal{P}_{\mathcal{A}}}; \tilde{U}_{\mathcal{A}}|Y) = 0$. Define $\Gamma_{\A} = I(X_{\mathcal{P}_{\mathcal{A}}}; \tilde{U}_{\mathcal{A}}|\tilde{U}_{\mathcal{A}^c}, Y)$. By the chain rule, 
$I(\tilde{U}_{\mathcal{A}}; \tilde{U}_{\mathcal{A}^c}|Y) - I(\tilde{U}_{\mathcal{A}}; \tilde{U}_{\mathcal{A}^c}|X_{\mathcal{P}_{\mathcal{A}}}, Y) = I(X_{\mathcal{P}_{\mathcal{A}}}; \tilde{U}_{\mathcal{A}}|Y) - \Gamma_{\mathcal{A}} = -\Gamma_{\mathcal{A}}$. Hence, the right-hand side of communication rates become $\Gamma_{\mathcal{A}} + [-\Gamma_{\mathcal{A}}]^- = 0$. Since $R_{\mathcal{A}} \ge 0$ for any non-empty subset $\A \subset \I$, the communication-rate constraints remain satisfied.
\item \emph{Step 4: Verification of the total sum rate of the communication rates and the leakage rate}. We have
\begin{align}
I(X_{\mathcal{P}}; \tilde{U}_{\mathcal{I}}|Y)
&\overset{\rm (a)}= I(X_{\mathcal{P}}; F, K_1, \dots, K_{\kappa-1}|Y)
\overset{\rm (b)}= I(X_{\mathcal{P}}; F|Y)
\overset{\rm (c)}= H(F|Y), \label{eq-94-eq}
\end{align}
where (a) holds because the mapping between $\tilde{U}_{\I}$ and $(F,K_1,\dots,K_{\kappa-1})$ is bijective, (b) holds because the masks are independent of the source variables, and (c) holds because $F$ is a function of $(X_{\mathcal{P}},Y)$. Similarly,
\begin{align}
I(X_{\mathcal{S}}; \tilde{U}_{\mathcal{I}}|Y)
= I(X_{\mathcal{S}}; F, K_1, \dots, K_{\kappa-1}|Y)
= I(X_{\mathcal{S}}; F|Y). \label{eq-95-eq}
\end{align}
For the total communication rate,
\begin{align*}
R_{\mathcal{I}}
\ge I(X_{\mathcal{P}}; U_{\mathcal{I}}|Y,Q)
\overset{\rm (a)}\ge I(X_{\mathcal{P}}; F|Y,Q)
\overset{\rm (b)}= H(F|Y)
\overset{\rm (c)}= I(X_{\mathcal{P}}; \tilde{U}_{\mathcal{I}}|Y),
\end{align*}
where (a) follows from the data processing inequality since the original auxiliary variables satisfy $H(F|U_{\mathcal{I}},Y,Q)=0$, (b) follows because $F$ is a function of $(X_{\mathcal{P}},Y)$ and $Q$ is independent of $(F,Y)$, and (c) follows from \eqref{eq-94-eq}. Likewise,
\begin{align*}
L
\ge I(X_{\mathcal{S}}; U_{\mathcal{I}}|Y,Q)
\overset{\rm (a)}\ge I(X_{\mathcal{S}}; F|Y,Q)
\overset{\rm (b)}= I(X_{\mathcal{S}}; F|Y)
\overset{\rm (c)}= I(X_{\mathcal{S}}; \tilde{U}_{\mathcal{I}}|Y),
\end{align*}
where (a) follows from the same data processing argument, (b) follows because $Q$ is independent of the sources, and (c) follows from \eqref{eq-95-eq}.
\end{itemize}
\item[(c)] By the support lemma \cite[Appx.~C]{GK}, preserving $P_{X_{\P_i}}$ requires $|\mathcal{X}_{\P_i}|-1$ elements with three more elements for $H(F|U_{\mathcal{I}},Y,Q)$, $I(X_{\mathcal{P}_i};U_i|Q)$, and $I(X_\S;U_\I|Q)$, yielding $|\mathcal{U}_i| \le |\mathcal{X}_{\mathcal{P}_i}|+2$.
\item[(d)] The bound on $|\mathcal{Q}|$ in Theorems~\ref{inner-bound}--\ref{th3-3-ll}  follows from the Fenchel--Eggleston--Carath\'eodory theorem \cite[Appx.~A]{GK}.
\end{enumerate}
\qed

\section{Proof of Theorem~\ref{inner-lemma}} \label{appx-cc-1}
The test channels $P_{U_i|X_{\Pi}}$ for $i \in \I$ are chosen such that there exists a reconstruction function $g^n(U^n_\I,Y^n) = \{g(U_{\I,t},Y_t)\}^n_{t=1}$ and $\mathbb{E}[d^n(f^n(X^n_{\P},Y^n),g^n(U^n_\I,Y^n))] \le D_{\max}$.

Since constraints \eqref{pri-def2} and \eqref{dis-def2} are preserved under time-sharing, the arguments in Appendix~\ref{appxc-a} extend to this setting, yielding the same conclusion as Lemma \ref{lemma-achiev}. That is, to prove the achievability of \eqref{ls-inner-eq}, it is sufficient to show the achievability of the corner point $((R_i)_{i \in \I},L,D)$ with
\begin{align}
R_i &= I(X_{\Pi};U_i|U_{[1:i-1]},Y) + 2\delta,~\forall i \in \I, \label{ri-eq-achi} \displaybreak[0] \\
L &= I(X_{\S};U_{\I}|Y) + \delta, \label{l-eq-achi} \displaybreak[0] \\
D &= \mathbb{E}\left[d(f(X_{\P},Y),g(U_{\I},Y))\right] + \delta. \label{7878-eqeq-achi}
\end{align}

\subsubsection{Random binning and common-randomness-aided coding scheme}\mbox{}
The random binning and coding schemes used in the proof for distributed lossless function computation in Appendix~\ref{appx-c-b} are employed in this proof.

\subsubsection{Coding Scheme Analysis}\mbox{}

\begin{Lemma}[\hspace{-0.01mm}{\cite[Lemma 5]{Yassaee-tit2014}}] \label{lm-9} Let $d: \mathcal{X} \times \mathcal{Y} \to [0, D_{\max}]$ be a bounded distortion measure. Suppose the joint distribution $P_{XY}$ satisfies $\mathbb{E}_{P_{XY}}[d(X,Y)] = D$. If $\Tilde{P}_{XY}$ is another joint distribution satisfying $\mathbb{V}(P_{XY}, \Tilde{P}_{XY}) < \epsilon$, then the expected distortion under $\Tilde{P}_{XY}$ satisfies $\mathbb{E}_{\Tilde{P}_{XY}}[d(X,Y)] \leq D + \epsilon D_{\max}$.
\end{Lemma}

\emph{Distortion}:~~~Let $\mathbb{E}_{\Tilde{P}^{b_\I}}[\cdot]$ denote the expectation under the law $\Tilde{P}^{b_\I}(\cdot)$ by the coding scheme. Observe that
\begin{align}
\E_{B_\I}\left[\mathbb{E}_{\Tilde{P}^{b_\I}}\left[d^n(f^n(X^n_{\P},Y^n),g^n(\hat{U}^n_{\I},Y^n))\right]\right] \nonumber \displaybreak[0] 
&\overset{\rm (a)}\le \E_{B_\I}\left[\mathbb{E}\Big[d^n(f^n(X^n_{\P},Y^n),g^n(U^n_{\I},Y^n))\Big]\right] + o(1) \nonumber \displaybreak[0] \\
&\overset{\rm (b)}= \E_{B_\I}\left[\mathbb{E}\Big[\frac{1}{n}\sum_{t=1}^n d(f_t(X_{\P,t},Y_t),g_t(U_{\I,t},Y_t))\Big]\right] + o(1) \nonumber \displaybreak[0] \\
&\overset{\rm (c)} \le \mathbb{E}\left[d(f(X_{\P},Y),g(U_{\I},Y))\right] + o(1) \nonumber \displaybreak[0] \\
&\overset{\rm (d)} \le D \label{d-eqls}
\end{align}
for large enough $n$, where (a) is due to \eqref{p-bar-1} and Lemma \ref{lm-9}, (b) holds from \eqref{distor-sum} as the sequences $(X^n_{\P},Y^n,U^n_{\I})$ generated from $P(\cdot)$ are i.i.d., (c) follows by the typical average lemma \cite[Ch.~2.4]{GK}, and (d) follows from $D$ defined in \eqref{7878-eqeq-achi}.

\emph{Communication and Leakage Rates}: Following \eqref{r-111} and \eqref{l-111}, the communication and leakage rates are bounded as follows:
\begin{align}
&\frac{1}{n}\log|\M_i| = R_i = I(X_{\P_i};U_i|U_{[1:i-1]},Y) + 2\delta, {\forall i \in \I}, \nonumber \displaybreak[0] \\
&{\E_{B_\I}}\left[\frac{1}{n}I_{\Tilde{P}^{B_\I}}(X^n_\S;M_\I|Y^n,C_\I)\right] \le L \label{l-eq-ls}
\end{align}
for sufficiently large $n$.

\subsubsection{Derandomization}\mbox{}
Similar to the lossless case, we define
\begin{align}
\delta'_n(b_\I,c_\I) &\triangleq \mathbb{P} \left[ \hat{U}_{\I}^n \neq U_{\I}^n \quad \text{or} \quad d^n(f^n(X_{\P}^n, Y^n), g^n(\hat{U}_{\I}^n, Y^n)) > D + \delta |(B_\I,C_\I)=(b_\I,c_\I) \right], \label{eqeq-95} \\
\ell'_n(b_\I,c_\I) &\triangleq \frac{1}{n} {I_{\tilde{P}^{b_\I}}} (X_{\S}^n; M_\I|Y^n,C_\I=c_\I). \label{eqeq-96}
\end{align}
for a given $(b_\I, c_\I)$. Equation \eqref{eqeq-95} captures the event of an incorrect SW decoding reconstruction or the resulting distortion greater than $D + \delta$, and \eqref{eqeq-96} measures the leakage rate.

{Under the i.i.d.~distribution $P^{b_\I}$, applying the weak law of large numbers yields the per-letter distortion measure (its mean is $D - \delta$ by \eqref{7878-eqeq-achi}), which gives
$\mathbb{P}\left[d^n(f^n(X^n_\mathcal{P}, Y^n), g^n(U^n_\mathcal{I}, Y^n)) > D + \delta\right] \longrightarrow 0~~\text{as}~~n \to \infty$. Hence, from \eqref{p-bar-1}, under the law $\Tilde{P}^{b_\I}$, it follows that
\begin{align}
\mathbb{P}\left[d^n(f^n(X^n_\mathcal{P}, Y^n), g^n(\hat{U}^n_\mathcal{I}, Y^n)) > D + \delta\right] \longrightarrow 0~~\text{as}~~n \to \infty. \label{wlln-eq}
\end{align}
Hence, by \eqref{p-bar-1} and \eqref{wlln-eq}}, it holds that  $\epsilon'_n \triangleq \mathbb{E}_{B_\I}\left[\mathbb{E}_{\tilde{P}^{B_\I}}[\delta'_n(B_\I,C_\I)]\right] \rightarrow 0~\text{as}~n \to \infty$, and $\mathbb{E}_{B_\I}\left[\mathbb{E}_{\tilde{P}^{B_\I}}[\ell'_n(B_\I, C_\I)]\right] \le L $ by \eqref{l-eq-ls}. Define $\mu'_n \triangleq \sqrt{\epsilon'_n} + n^{-1}$ ($\mu'_n \to 0$ and $\frac{\epsilon'_n}{\mu'_n} \to 0$ as $n \to \infty$). By Markov's inequality, we have
\[
\mathbb{P}\left\{ \delta'_n(B_\I,C_\I) \le \mu'_n, \ell'_n(B_\I, C_\I) \le L + \delta \right\} \ge 1 - \frac{\epsilon'_n}{\mu'_n} - \frac{L}{L + \delta} = \frac{\delta}{L + \delta} - \frac{\epsilon'_n}{\mu'_n} \overset{n \to \infty}\longrightarrow \frac{\delta}{L + \delta} > 0.
\]
The strict positivity of this probability guarantees the existence of a realization pair $(b_\I, c_\I)$ satisfying
\begin{align}
\mathbb{P} \left[d^n(f^n(X_{\P}^n, Y^n), g^n(\hat{U}_{\I}^n, Y^n)) > D + \delta | (B_\I,C_\I) = (b_\I,c_\I) \right] &\le \mu'_n, \label{eq-mu-n}\\
\frac{1}{n}I_{\Tilde{P}^{b_\I}}(X^n_\S;M_\I|Y^n,C_\I=c_\I) &\le L + \delta. \nonumber
\end{align}
Under the fixed pair $(b_\I, c_\I)$, we define a random variable $D_n = d^n(f^n(X_{\P}^n, Y^n), g^n(\hat{U}_{\I}^n, Y^n))$ and an event $\mathcal{D}_n = \{(X^n_{\P},Y^n,U_{\I}^n,\hat{U}_{\I}^n): D_n > D + \delta\}$. Then, we have
\begin{align*}
&\mathbb{E}_{\Tilde{P}^{b_\I}}\left[d^n(f^n(X^n_{\P},Y^n),g^n(\hat{U}^n_{\I},Y^n))| (B_\I,C_\I) = (b_\I,c_\I) \right] \nonumber \displaybreak[0] \\
&=\mathbb{E}_{\Tilde{P}^{b_\I}}\left[D_n \cdot\mathds{1}[\mathcal{D}^c_n] | (B_\I,C_\I) = (b_\I,c_\I) \right] +\mathbb{E}_{\Tilde{P}^{b_\I}}\left[D_n\cdot\mathds{1}[\mathcal{D}_n] | (B_\I,C_\I) = (b_\I,c_\I) \right] \nonumber \displaybreak[0] \\
&\overset{\rm (a)}\le (D + \delta)\mathbb{P}\left[\mathcal{D}^c_n | (B_\I,C_\I) = (b_\I,c_\I) \right] + D_{\rm max}\mathbb{P}\left[\mathcal{D}_n | (B_\I,C_\I) = (b_\I,c_\I) \right] \nonumber \displaybreak[0] \\
&\overset{\rm (b)}\le D + \delta + D_{\max}\mu'_n,
\end{align*}
where (a) holds because $D_n \le D + \delta$ for $\mathds{1}[\mathcal{D}^c_n] = 1$ and $D_n \le D_{\rm max}$, and (b) holds by \eqref{eq-mu-n} and because $\mathbb{P}\left[\mathcal{D}^c_n | (B_\I,C_\I) = (b_\I,c_\I) \right] \le 1$.   By taking $n \to \infty$ and $\delta \downarrow 0$, the corner point $((R_i)_{i \in \mathcal{I}}, L, D)$ specified in \eqref{ri-eq-achi}--\eqref{7878-eqeq-achi} is achieved.
\qed

\section{Proof of Theorem~\ref{outer-lemma}} \label{appx-cc-2}

Assume that a rate tuple $((R_i)_{i \in \I},L,D)$ is achievable according to Definition \ref{def2}, and define {$U_{i,t}\triangleq (M_i,X^{t-1}_{\S_i})$ and $Q_t \triangleq Y^{n \backslash t}$, which satisfies $(Q_t,U_{i,t})-X_{\Pi,t}-(X_{\P\setminus\Pi,t}X_{\S,t},Y_t)$ for $i \in \I$ as shown in \eqref{mk-converse}}. Since the analyses of communication and leakage rates are similar to those in the lossless case, the details are omitted, and only the analysis of the distortion is provided. From \eqref{dis-def2}, we have
\begin{align}
    D + \delta &\ge \mathbb{E}\left[d^n(f^n(X^n_{\P},Y^n),\psi^n(M_{\I},Y^n))\right] \nonumber \displaybreak[0] \\
    &= \mathbb{E}\left[\frac{1}{n}\sum_{t=1}^n d(f_t(X_{\P,t},Y_t),\psi_t(M_{\I},Y^n))\right] \nonumber \displaybreak[0] \\
    &\overset{\rm (a)}= \frac{1}{n}\sum_{t=1}^n\mathbb{E}\left[d(f_t(X_{\P,t},Y_t),g_t(U_{\I,t},Y_t,Q_t))\right] \nonumber \displaybreak[0] \\
    &\overset{\rm (b)}= \frac{1}{n}\sum_{t=1}^n\mathbb{E}\left[d(f_T(X_{\P,T},Y_T),g_T(U_{\I,T},Y_T,Q_T))\mid T=t\right] \nonumber \displaybreak[0] \\
    &= \mathbb{E}\left[d(f_T(X_{\P,T},Y_T),g_T(U_{\I,T},Y_T,Q_T))\right] \nonumber \displaybreak[0] \\
    &\overset{\rm (c)}= \mathbb{E}\left[d(f(X_{\P},Y),g(U_{\I},Y,Q))\right], \label{d-ana}
\end{align}
where (a) holds by selecting a reconstruction function
$g_t(U_{\I,t},Y_t,Q_t)=\psi_t(M_{\I},Y^n)$ for all
$t\in[1:n]$, which is possible because $(U_{\I,t},Y_t,Q_t)$
contains $(M_{\I},Y^n)$, (b) holds by introducing a random
variable $T$ uniformly distributed over $[1:n]$ and independent
of all previously defined random variables, and (c) holds by
defining $Q\triangleq(Q_T,T)$, $U_i\triangleq U_{i,T}$,
$X_{\Pi}\triangleq X_{\Pi,T}$, $X_{\Si}\triangleq X_{\Si,T}$,
$Y\triangleq Y_T$, and
$g(u_{\I},y,(q,t))\triangleq g_t(u_{\I},y,q)$, and using
$f_t=f$ for every $t$.
Moreover, \eqref{mk-converse} and the i.i.d. source assumption
imply that $(Q,U_i)-X_{\Pi}-(X_{\P\setminus\Pi},X_{\S},Y)$
for all $i\in\I$, and that $Q$ is independent of
$(X_{\P},X_{\S},Y)$.
The cardinality bound of  $U_i$ is shown in Lemma \ref{CB-ls}. Finally, by letting $n \rightarrow \infty$ and $\delta \downarrow 0$, any achievable rate tuple is included in \eqref{ls-th-eq}.
\qed

\section{Proof of Lemma \ref{CB-ls}} \label{Appx-CB-ls}
The proofs of (a), (c), and (d) are the same as those in Lemma \ref{CB-ll} (cf.~Appendix~\ref{Appx-CB-ll}), so we provide only the proof of (b). Note that the reconstruction function $\hat{F} = g(U_\I,Y,Q)$ takes values from the set $\mathcal{F}$. Similar to the proof of Lemma \ref{CB-ll}, we consider $\mathcal{F} = [0:m-1]$ and choose mutually independent uniform masks $K_1, \dots, K_{\kappa-1}$ on $\mathcal{F}$, independent of source variables, including $F$ and $\hat{F}$. Define random masks
$\tilde{U}_i = K_i~(i \in [1:\kappa-1]),$ and $\tilde{U}_{\kappa} = \hat{F} - \sum_{i=1}^{\kappa-1} K_i \pmod{m}$.

By the same arguments as in Steps 2 and 3 of the proof in Lemma \ref{CB-ll}, it follows that the replacement auxiliary variables $\tilde{U}_\A$ for $\A \subset \I$ are uniform,
the Markov conditions in \eqref{mk-ls} hold, and the communication rates with non-empty subsets in \eqref{ls-th-eq} remain satisfied. {Then, define a new decoder by
$\tilde{g}(\tilde{U}_{\mathcal{I}}, Y) = \sum_{i=1}^{\kappa} \tilde{U}_i \pmod{m}$. Its output equals the original reconstruction $\hat{F}$, consequently,
\begin{align}
\mathbb{E}\left[d(F, \tilde{g}(\tilde{U}_{\mathcal{I}}, Y))\right] = \mathbb{E}[d(F, \hat{F})] \le D. \label{eq-110-eq}
\end{align}
For either a random variable $W = X_{\P}$ or $W = X_{\S}$, we have
\begin{align}
I(W; \tilde{U}_{\I}|Y) \overset{\rm (a)}= I(W; \hat{F}|Y)
\overset{\rm (b)}\le I(W; U_{\I}, Q|Y) 
\overset{\rm (c)}= I(W; U_{\I}|Y, Q), \label{eq-111-eq}
\end{align}
where (a) follows from the independence among the masks, (b) holds because   $\hat{F}$ is a function of $(U_{\mathcal{I}}, Y, Q)$, and (c) is due to the independence of $Q$ from $(W, Y)$}.

  {Equation \eqref{eq-110-eq} implies that the distortion constraint is preserved under $\tilde{U}_{\mathcal{I}}$ as the resulting distortion does not increase, while \eqref{eq-111-eq} ensures that the total sum rate of the communication rates and the leakage rate are  preserved. Therefore, replacing $U_{\mathcal{I}}$ with $\tilde{U}_{\mathcal{I}}$ preserves all conditions in Theorem \ref{outer-lemma}. Since each $\tilde{U}_i$ for $i \in \I$ takes values in $[0:m-1]$, this establishes that it suffices to restrict $|\tilde{\mathcal{U}}_i| \le |\mathcal{F}|$}.
\qed

\bibliographystyle{IEEEtran}
\bibliography{references}

@ARTICLE{yamamoto-1983,
  author={Yamamoto, H.},
  journal={IEEE Trans. Inf. Theory}, 
  title={A source coding problem for sources with additional outputs to keep secret from the receiver or wiretappers (Corresp.)}, 
  year={1983},
  volume={29},
  number={6},
  pages={918-923},
  month = nov
}

@ARTICLE{Sankar-2013,
  author={Sankar, Lalitha and Rajagopalan, S. Raj and Poor, H. Vincent},
  journal={IEEE Trans. Inf. Forensics Secur.}, 
  title={Utility-Privacy Tradeoffs in Databases: An Information-Theoretic Approach}, 
  year={2013},
  volume={8},
  number={6},
  pages={838-852},
  month=mar
}

@ARTICLE{deepesh-2016,
  author={Data, Deepesh and Prabhakaran, Vinod M. and Prabhakaran, Manoj M.},
  journal={IEEE Trans. Inf. Theory}, 
  title={Communication and Randomness Lower Bounds for Secure Computation}, 
  year={2016},
  volume={62},
  number={7},
  pages={3901-3929},
  month = jul
}

@article{chou2014separation,
  title={Separation of reliability and secrecy in rate-limited secret-key generation},
  author={Chou, Remi A and Bloch, Matthieu R},
  journal={IEEE Trans. Inf. Theory},
  volume={60},
  number={8},
  pages={4941--4957},
  month = aug,
  year={2014}}

@ARTICLE{deepesh-2020,
  author={Data, Deepesh and Kurri, Gowtham R. and Ravi, Jithin and Prabhakaran, Vinod M.},
  journal={IEEE Trans. Inf. Theory}, 
  title={Interactive Secure Function Computation}, 
  year={2020},
  volume={66},
  number={9},
  pages={5492-5521},
  month = sep
}

@ARTICLE{naghibis2015,
  author={Naghibi, Farshad and Salimi, Somayeh and Skoglund, Mikael},
  journal={IEEE Trans. Inf. Forensics Secur.}, 
  title={The {CEO} Problem With Secrecy Constraints}, 
  year={2015},
  volume={10},
  number={6},
  pages={1234-1249}
}

@ARTICLE{km-1979,
  author={Korner, J. and Marton, K.},
  journal={IEEE Trans. Inf. Theory}, 
  title={How to encode the modulo-two sum of binary sources (Corresp.)}, 
  year={1979},
  volume={25},
  number={2},
  pages={219-221},
  month = mar
}

@ARTICLE{ah-1983,
  author={Ahlswede, R. and Te Sun Han},
  journal={IEEE Trans. Inf. Theory}, 
  title={On source coding with side information via a multiple-access channel and related problems in multi-user information theory}, 
  year={1983},
  volume={29},
  number={3},
  pages={396-412},
  month = may
}

@INPROCEEDINGS{chandra-2020,
  author={Nair, Chandra and Wang, Yan Nan},
  booktitle={Proc. IEEE Int. Symp. Inf. Theory (ISIT)}, 
  title={On optimal weighted-sum rates for the modulo sum problem}, 
  year={2020},
  volume={},
  address={Los Angeles, CA, USA},
  pages={2416-2420},
  month = jun}

@INPROCEEDINGS{milad-2020,
  author={Sefidgaran, Milad and Tchamkerten, Aslan},
  booktitle={Proc. IEEE Inf. Theory Workshop (ITW)}, 
  title={Zero-Error Sum Modulo Two with a Common Observation}, 
  year={2021},
  volume={},
  address={Riva del Garda, Italy},
  pages={1-5},
  month= apr
}

@INPROCEEDINGS{tharindu-2022,
  author={Adikari, Tharindu and Draper, Stark},
  booktitle={Proc. IEEE Int. Symp. Inf. Theory (ISIT)}, 
  title={Two-terminal source coding with common sum reconstruction}, 
  year={2022},
  volume={},
  address={Espoo, Finland},
  pages={1420-1424},
  month = {Jun./Jul.}
}

@ARTICLE{va-tifs-2025,
  author={Yachongka, Vamoua and Yagi, Hideki and Ochiai, Hideki},
  journal={IEEE Trans. Inf. Forensics Secur.}, 
  title={Outer Bounds on the {CEO} Problem With Privacy Constraints}, 
  year={2025},
  volume={20},
  number={},
  pages={1566-1581},
  month = feb
}

@ARTICLE{vamoua-tifs,
  author={Yachongka, Vamoua and Yagi, Hideki and Ochiai, Hideki},
  journal={IEEE Trans. Inf. Forensics Secur.}, 
  title={Key Agreement Using Physical Identifiers for Degraded and Less Noisy Authentication Channels}, 
  year={2023},
  volume={18},
  number={},
  pages={5316-5331},
  month = aug
}

@INPROCEEDINGS{vr-isit2025,
  author = {Yachongka, Vamoua and Chou, Rémi A.},
  booktitle={Proc. IEEE Int. Symp. Inf. Theory (ISIT)}, 
  title = {Privacy-Constrained Lossy Function Computation in Distributed Databases},
  address={Ann Arbor, Michigan, USA},
  month=jun,
  year = {2025}
}

@ARTICLE{Ali-2021,
  author={Ghubaish, Ali and Salman, Tara and Zolanvari, Maede and Unal, Devrim and Al-Ali, Abdulla and Jain, Raj},
  journal={IEEE Internet Things J.}, 
  title={Recent Advances in the {I}nternet-of-{M}edical-{T}hings ({I}o{MT}) Systems Security}, 
  year={2021},
  volume={8},
  number={11},
  pages={8707-8718},
  month = jun
}

@ARTICLE{wata-2020,
  author={Watanabe, Shun},
  journal={IEEE Trans. Inf. Theory}, 
  title={A Classification of Functions in Multiterminal Distributed Computing}, 
  year={2020},
  volume={66},
  number={10},
  pages={6169-6183},
  month = oct
}

@ARTICLE{alphan-2023,
  author={Şahin, Alphan and Yang, Rui},
  journal={IEEE Commun. Surv. Tutor.}, 
  title={A Survey on Over-the-Air Computation}, 
  year={2023},
  volume={25},
  number={3},
  pages={1877-1908}
}

@ARTICLE{chou-isit2024,
  author={Chou, R\'emi A. and Kliewer, J\"org and Yener, Aylin},
  journal={IEEE Trans. Inf. Theory}, 
  title={Private Sum Computation: Trade-Offs Between Communication, Randomness, and Privacy}, 
  year={2026},
  month = apr,
  volume={72},
  number={4},
  pages={2472-2482}
}

@Article{gunlu-en-d,
AUTHOR = {Günlü, Onur},
TITLE = {Function Computation under Privacy, Secrecy, Distortion, and Communication Constraints},
JOURNAL = {Entropy},
VOLUME = {24},
YEAR = {2022},
NUMBER = {1},
ARTICLE-NUMBER = {110},
month = jan
}

@book{gallager-2013,
  author         = {R. G. Gallager},
  publisher      = {Cambridge Univ. Press},
  address        = {Cambridge, U.K.},
  title          = {Stochastic Processes: Theory for Applications},
  year           = {2013}
}

@ARTICLE{chou2022,
  author={Chou, Rémi A. and Kliewer, Joerg},
  journal={IEEE Trans. Inf. Theory}, 
  title={Function Computation Without Secure Links: Information and Leakage Rates}, 
  year={2025},
  volume={71},
  number={11},
  pages={9004-9018},
  month=nov
}

@ARTICLE{villard-2013,
  author={Villard, Joffrey and Piantanida, Pablo},
  journal={IEEE Trans. Inf. Theory}, 
  title={Secure Multiterminal Source Coding With Side Information at the Eavesdropper}, 
  year={2013},
  volume={59},
  number={6},
  pages={3668-3692},
  month=jun
}

@ARTICLE{amirraza-2024,
  author={Zamani, Amirreza and Oechtering, Tobias J. and Skoglund, Mikael},
  journal={IEEE Trans. Inf. Theory}, 
  title={On the Privacy-Utility Trade-Off With and Without Direct Access to the Private Data}, 
  year={2024},
  volume={70},
  number={3},
  pages={2177-2200},
  month = mar
}

@ARTICLE{tu2019,
  author={Tu, Wenwen and Lai, Lifeng},
  journal={IEEE Trans. Inf. Theory}, 
  title={On Function Computation With Privacy and Secrecy Constraints}, 
  year={2019},
  month = jun,
  volume={65},
  number={10},
  pages={6716-6733}
}

@ARTICLE{Yamamoto-fc,
  author={Yamamoto, H.},
  journal={IEEE Trans. Inf. Theory}, 
  title={Wyner - {Z}iv theory for a general function of the correlated sources (Corresp.)}, 
  year={1982},
  volume={28},
  number={5},
  pages={803-807},
  month=sep
}

@ARTICLE{guang-2024,
  author={Guang, Xuan and Zhang, Ruze},
  journal={IEEE Trans. Inf. Theory}, 
  title={Zero-Error Distributed Compression of Binary Arithmetic Sum}, 
  year={2024},
  volume={70},
  number={5},
  pages={3100-3117},
  month=may
}

@ARTICLE{zhang-2024,
  author={Zhang, Ruze and Guang, Xuan and Yang, Shenghao and Niu, Xueyan and Bai, Bo},
  journal={IEEE J. Sel. Areas Inf. Theory}, 
  title={Computation of Binary Arithmetic Sum Over an Asymmetric Diamond Network}, 
  year={2024},
  volume={5},
  number={},
  pages={585-596},
  month=sep
}

@ARTICLE{goldenbuam,
  author={Goldenbaum, Mario and Boche, Holger and Stańczak, Sławomir},
  journal={IEEE Trans. Wireless Commun.}, 
  title={Nomographic Functions: Efficient Computation in Clustered {G}aussian Sensor Networks}, 
  year={2015},
  volume={14},
  number={4},
  pages={2093-2105},
  month=apr
}

@ARTICLE{nazer-2007,
  author={Nazer, Bobak and Gastpar, Michael},
  journal={IEEE Trans. Inf. Theory}, 
  title={Computation Over Multiple-Access Channels}, 
  year={2007},
  volume={53},
  number={10},
  pages={3498-3516},
  month = oct
}

@ARTICLE{frey-2025,
  author={Frey, Matthias and Bjelaković, Igor and Gastpar, Michael C. and Zhu, Jingge},
  journal={IEEE Trans. Inf. Theory}, 
  title={Simultaneous Computation and Communication Over {MAC}}, 
  year={2025},
  volume={71},
  number={5},
  pages={3644-3665},
  month = may
}

@ARTICLE{Jeon-2014,
  author={Jeon, Sang-Woon and Wang, Chien-Yi and Gastpar, Michael},
  journal={IEEE Trans. Inf. Theory}, 
  title={Computation Over {G}aussian Networks With Orthogonal Components}, 
  year={2014},
  volume={60},
  number={12},
  pages={7841-7861},
  month=oct
}

@ARTICLE{orlitsky-2001,
  author={Orlitsky, A. and Roche, J.R.},
  journal={IEEE Trans. Inf. Theory}, 
  title={Coding for computing}, 
  year={2001},
  month = aug,
  volume={47},
  number={3},
  pages={903-917}
}

@inproceedings{va-remi-isit2026,
  author          = {Yachongka, Vamoua
                 and Chou, Rémi A.},
  title           = {Privacy-utility trade-offs for private function computation in databases},
  booktitle       = {Proc. IEEE Int. Symp. Inf. Theory (ISIT)},
  address         = {Guangzhou, China},
  year            = {2026},
  month           = {Jun./Jul.}
}

@ARTICLE{kitti-et-al2016,
  author={Kittichokechai, Kittipong and Chia, Yeow-Khiang and Oechtering, Tobias J. and Skoglund, Mikael and Weissman, Tsachy},
  journal={IEEE Trans. Inf. Theory}, 
  title={Secure Source Coding With a Public Helper}, 
  year={2016},
  volume={62},
  number={7},
  pages={3930-3949},
  month = Jul
  }

@ARTICLE{kuzuoka-2015,
  author={Kuzuoka, Shigeaki and Watanabe, Shun},
  journal={IEEE Trans. Inf. Theory}, 
  title={A Dichotomy of Functions in Distributed Coding: An Information Spectral Approach}, 
  year={2015},
  volume={61},
  number={9},
  pages={5028-5041},
  month = sep
}

@ARTICLE{milad2016,
  author={Sefidgaran, Milad and Tchamkerten, Aslan},
  journal={IEEE Trans. Inf. Theory}, 
  title={Distributed Function Computation Over a Rooted Directed Tree}, 
  year={2016},
  month = dec,
  volume={62},
  number={12},
  pages={7135-7152}
}

@ARTICLE{ekrem-2014,
  author={Ekrem, Ersen and Ulukus, Sennur},
  journal={IEEE Trans. Inf. Theory}, 
  title={The Secrecy Capacity Region of the {G}aussian {MIMO} Multi-Receiver {W}iretap Channel}, 
  year={2011},
  volume={57},
  number={4},
  pages={2083-2114},
  month = apr
}

@ARTICLE{hassan-2024,
  author={ZivariFard, Hassan and Chou, Rémi A.},
  journal={IEEE J. Sel. Areas Inf. Theory}, 
  title={Secure Source Coding Resilient Against Compromised Users via an Access Structure}, 
  year={2024},
  volume={5},
  pages={478-492},
  month = jun
}

@ARTICLE{Ekrem-2013,
  author={Ekrem, Ersen and Ulukus, Sennur},
  journal={IEEE Trans. Inf. Theory}, 
  title={Secure Lossy Transmission of Vector {G}aussian Sources}, 
  year={2013},
  volume={59},
  number={9},
  pages={5466-5487},
  month= sep
  }

@Inbook{Edmonds2003,
author="Edmonds, Jack",
title="Submodular Functions, Matroids, and Certain Polyhedra",
year="2003",
publisher="Springer Berlin Heidelberg",
address="Berlin, Heidelberg",
pages="11--26",
month = jan
}

@ARTICLE{wyner-ziv,
  author={Wyner, A. and Ziv, J.},
  journal={IEEE Trans. Inf. Theory}, 
  title={The rate-distortion function for source coding with side information at the decoder}, 
  year={1976},
  volume={22},
  number={1},
  pages={1-10},
  month= jan
}

@ARTICLE{va-remi-tifs,
  author={Yachongka, Vamoua and Chou, Rémi A.},
  journal={IEEE Trans. Inf. Forensics Secur.}, 
  title={Secret-Key Generation From Private Identifiers Under Channel Uncertainty}, 
  year={2025},
  volume={20},
  number={},
  pages={12141-12156},
  month=oct
}

@book{Domdouzis2021,
  author    = {Konstantinos Domdouzis and Peter Lake and Paul Crowther},
  title     = {Concise Guide to Databases: A Practical Introduction},
  publisher = {Springer},
  address   = {Cham},
  year      = {2021}
}

@ARTICLE{kuzuoka-2017,
  author={Kuzuoka, Shigeaki and Watanabe, Shun},
  journal={IEEE Trans. Inf. Theory}, 
  title={On Distributed Computing for Functions With Certain Structures}, 
  year={2017},
  volume={63},
  number={11},
  pages={7003-7017},
  month= nov
}

@article{gunlu-et-al2018,
  author          = {G\"{u}nl\"{u}, Onur
                 and Kittichokechai, Kittipong
                 and Schaefer, Rafael F.
                 and Caire, Giuseppe},
  title           = {Controllable identifier measurements for private authentication with secret keys},
  journal         = {IEEE Trans. Inf. {Forensics} Secur.},
  volume          = {13},
  number          = {8},
  pages           = {1945-1959},
  year            = {2018},
  month           = aug
}

@INPROCEEDINGS{lu-2022,
  author={Lu, Jian and Xu, Yinfei and Shao, Shuo},
  booktitle={Proc. IEEE Inf. Theory Workshop (ITW)}, 
  title={Functional Privacy for Distributed Function Computation}, 
  year={2022},
  address={Mumbai, India},
  pages={1-6},
  month = nov
}

@article{csiszar1996,
  title={Almost Independence and Secrecy Capacity},
  author={Csisz{\'a}r, Imre},
  journal={Probl. Inf. Transm.},
  volume={32},
  number={1},
  pages={40--47},
  year={1996},
  month = jan
}

@ARTICLE{Yassaee-tit2014,
  author={Yassaee, Mohammad Hossein and Aref, Mohammad Reza and Gohari, Amin},
  journal={IEEE Trans. Inf. Theory}, 
  title={Achievability Proof via Output Statistics of Random Binning}, 
  year={2014},
  volume={60},
  number={11},
  pages={6760-6786},
  month = nov
}

@book{cover,
author = {Cover, Thomas M. and Thomas, Joy A.},
title = {Elements of Information Theory },
year = {2006},
edition = {2},
publisher = {John Wiley \& Sons},
address = {NJ, USA}
}

@article{bergmans1974,
  author          = {Bergmans, P.},
  title           = {A simple converse for broadcast channels with additive white {G}aussian noise {(Corresp.)}},
  journal         = {IEEE Trans. Inf. Theory},
  volume          = {20},
  number          = {2},
  pages           = {279-280},
  year            = {1974},
  month           = mar
}

@ARTICLE{rana2022,
  author={Rana, Vidhi and Chou, Rémi and Kwon, Hyuck M.},
  journal={IEEE Trans. Inf. Theory}, 
  title={Information-Theoretic Secret Sharing From Correlated {G}aussian Random Variables and Public Communication}, 
  year={2022},
  volume={68},
  number={1},
  pages={549-559},
  month = jan
}

@book{GK,
  author         = {El Gamal, Abbas
                and Kim, Young-Han},
  publisher      = {Cambridge Univ. Press},
  address        = {Cambridge, U.K.},
  title          = {Network Information Theory},
  year           = {2011}
}

@article{milad-arxiv,
  author       = {Milad Sefidgaran and
                  Aslan Tchamkerten},
  title        = {Computing a Function of Correlated Sources},
  journal      = {arXiv:1107.5806},
  year         = {2012},
  month         = oct
}

\end{document}